\documentclass[screen,acmsmall,authorversion,nonacm]{acmart}

\setcopyright{cc}

\begin{document}
\title{Navigating through CS1: The Role of Self-Regulation and Supervision in Student Progress}

\author{Ville Isomöttönen}
\orcid{0000-0002-5274-236X}
\affiliation{%
  \institution{University of Jyväskylä}
  \department{Faculty of Information Technology}
  \city{Jyväskylä}
  \country{Finland}
}
\email{ville.isomottonen@jyu.fi}

\author{Denis Zhidkikh}
\orcid{0000-0003-1335-8346}
\affiliation{%
  \institution{University of Jyväskylä}
  \department{Faculty of Information Technology}
  \city{Jyväskylä}
  \country{Finland}
}

\authorsaddresses{Corresponding author: Ville Isomöttönen,
\href{mailto:ville.isomottonen@jyu.fi}{ville.isomottonen@jyu.fi}; University of Jyväskylä, Faculty of Information Technology, P.O. Box 35, FI-40014 Jyväskylä, Finland}

\begin{abstract}
The need for students' self-regulation for fluent transitioning to university studies is known. Our aim was to
integrate study-supportive activities with course supervision activities within CS1. We educated TAs to pay attention to students' study ability and self-regulation. An interview study ($N=14$) was undertaken to investigate this approach. A thematic analysis yielded rather mixed results in light of our aims. Self-regulation was underpinned by the influences external to our setting, including labor market-related needs, earlier crises in study habits, and personal characteristics such as passion, grit, creativity, and valuation of utility. Safety in one-to-one supervision was considered essential, while shyness, fear, and even altruism caused self-handicapping during the course. Students were aware of their learning styles and need for self-regulation, while did not always know how to self-regulate or preferred to externalize it. The results highlight that supporting self-regulation should be integrated with students' personal histories and experiences, and thereby calls attention to transformative learning pedagogies. The thematization can help to understand CS1 students' self-regulation processes and improve CS1 support practices.
\end{abstract}



\maketitle

\section{Introduction}

New university students face numerous challenges, including managing increased workloads, adapting to new learning styles, and taking greater responsibility for their learning \citep{Perry2001AcademicControlAction, Meehan2018SearchFeelingBelonging}.
In Computer Science (CS) education, %
challenges related to self-regulation (SRL), self-efficacy, and help-seeking during the first year have been reported in many studies \citep[e.g.,][]{Hamalainen2019, Kinnunen2006WhyStudentsDrop, Petersen2016RevisitingWhyStudents}.
Consequently, universities have fostered students' integration through various services and intervention programs.
For instance, universities provide professional counseling, where students consult faculty members for assistance with learning, administrative issues, and career guidance \citep[e.g.,][]{ThiVanPham2021StudentSupportServices}. %
Additionally, reduced attrition rates have been reported after implementing first-year introductory modules that teach basic meta-learning skills such as information seeking, academic writing, and SRL skills \citep[e.g.,][]{Turner2017EasingTransitionFirst, Perin2011FacilitatingStudentLearning}.
Similar subject-oriented ``transition courses'' that prepare students for core curriculum content have also been found to improve retention and promote social integration \citep[e.g.,][]{Kangas2017HowFacilitateFreshmen}.

Furthermore, Teaching Assistant (TA) programs that employ undergraduate or graduate students as course lecturers' assistants and a peer support resource have proven popular in CS1 \citep{Mirza2019UndergraduateTeachingAssistants}.
TAs provide course-specific technical assistance to students, typically through workshops or personal tutoring \citep[e.g.,][]{Minnes2018LightweightTechniquesSupport}.
They also help to alleviate the workload of the course lecturers by editing course materials \citep{vanDam2018ReflectionsIntroductoryCS}, grading assignments \citep{Minnes2018LightweightTechniquesSupport}, and holding office hours.

While students are provided with various support avenues, current practices tend to separate student integration from the course context. For instance, organizational support like professional study counselors and introductory modules teach SRL skills but are generally situated outside any courses \citep{Mason2019EvaluationStudySkills}. %
Students may also be reluctant to seek external support if they hold negative attitudes or stigma towards them \citep[e.g.,][]{Giovazolias2010AssessmentGreekUniversity}.
In that sense, TA programs may help bring counseling and comprehensive support ``closer to the student''.
In a comprehensive review of TA use in CS, \citet{Mirza2019UndergraduateTeachingAssistants} conclude that most TA programs focus primarily on training TAs to provide technical, course-specific support and general instruction skills.
Moreover, near-peer mentoring programs have shown promise as a means of course-integrated student counseling support \citep[e.g.,][]{PonEtAl17}. 
The cited study demonstrates an approach that was favorably received by the students according to the feedback. However, we believe that more detailed research is needed to further elaborate on different scenarios that explain students' self-regulatory processes and use of supervision in such settings. 

The present study initially aimed to examine the effects of training TAs to support students' studying, and not only progress with assignments, during CS1. 
The TAs were educated regarding the struggles typically faced by CS1 students and how to approach them. During the course, students participated in periodic one-on-one supervision sessions with TAs, where they could privately discuss studying and seek solutions to learning problems. Students were frequently asked to write reflections about their progress in a virtual learning environment for TAs to view.
Participated students were invited for personal one-on-one interviews after the course to evaluate the TA support. 
Altogether 14 students were interviewed. 
During the interviews, we noticed that students' perception of the CS1 course and the one-to-one supervision sessions was clearly coupled with influences arising from students' personal histories and experiences -- and not from the study arrangement.
As such, we repositioned our research from merely evaluating the offered one-to-one sessions to illuminating self-regulation and help-seeking more broadly in CS1 context. We performed a qualitative analysis seeking to answer the question of \emph{how self-regulatory processes and use of supervision emerged during CS1}. We conjectured that this goal usefully complements the studies focusing on TAs role and their training, such as one by \citet{PonEtAl17}.
The study follows an inductive approach in which the resultant thematization provides the answer to the set question. 

\section{Research background} \label{sec:theory-and-related}

We refer to two theoretical frameworks: SRL, which scoped the inductive study, and transformative learning, which appeared to emerge in our results and is hence addressed.

SRL refers to a student's ability to approach learning by setting goals and proactively working towards them while monitoring their progress and adjusting their learning methods as necessary \citep{Schunk2005}.
In SRL, students set goals, use various learning resources and learning strategies, monitor their progress, and reflect on their performance
and outcomes to refine their future learning actions \citep{Zimmerman2008InvestigatingSelfRegulationMotivation}.
Self-regulation manifests in learning by a student regulating various cognitive aspects, such as motivation, self-efficacy beliefs, behavior, use of time, and the learning context \citep[e.g.,][]{Pintrich2000RoleGoalOrientation}.
Self-regulation skills are closely linked to metacognition and self-efficacy, which are actively researched in CS education \citep{Steinhorst2020RevisitingSelfEfficacyIntroductory, Prather2020WhatWeThinka}.

We identify at least four perspectives of SRL in computing education research: students' approach to studying,  problem-solving process, specific SRL aspects such as help-seeking, and SRL interventions.  The study by Falkner et al. \citep*{Falkner2014IdentifyingComputerScience} exemplifies the first two aspects, identifying various SRL strategies used by CS students. The results included general planning and time management strategies, visualization, problem decomposition, and knowledge building via practice. \citet {Loksa2016RoleSelfRegulationProgramming} primarily focused on the process of SRL when solving programming problems and noticed that SRL strategies proved more successful when a student had sufficient programming knowledge for the task. The help-seeking perspective is present in the study by \citet{Doebling2021PatternsAcademicHelpSeeking}, who surveyed CS students about their approaches and found that CS students most often seek help from online and peers before turning to the course teachers. Interventions were summarized by \citet{Prather2020WhatWeThinka}, including different types of reflection tasks, scaffolding for problem-solving and tactics to raise students' awareness of their habits. The review disclosed mixed results with the interventions.

Generally, Loksa and Ko \citep*{Loksa2016RoleSelfRegulationProgramming} highlighted that self-regulation among first-year CS students is often infrequent and at a surface level. They further concluded that SRL lacked consistency and discipline. Lack of regulatory skills among introductory programming students was also highlighted by \citet{FerEtAl23}. Conforming to the previous research, \citet{Loksa2016RoleSelfRegulationProgramming} also observed signs of improvement from CS1 to CS2, potentially arising from indirect or independent learning of SRL. In this connection, it is interesting to acknowledge studies in which students were provided with the flexibility to set their goals and yet slipped from their goals. This occurred in a functional programming (FP) course taken after CS1 and CS2  \citep{IsoTir16}, while \citet{CamEtAl19} reported similar difficulties with the flexibility provided on CS1. %

Regarding TLT, \citet{Dir98} helps us to note several perspectives. He identified adult learning as being based on the adaptation to the changing social-cultural surroundings or, on the other hand, self-reflection. This latter scenario alluded to Transformative Learning Theory (TLT), where \citet{Dir98} identified several perspectives. \citet{Dal86} had emphasized the developmental aspect through the reworking of meaning structures, while \citet{Boy91} had focused on depth psychology and Jungian individuation. 
The one of particular interest to us is the widely known theorization typically traced back to Mezrow's notion of perspective transformation \citep{Mez78}. %
\citet{Mez78} acknowledged continuing learning via lived experience, and the perspective transformation referred to the anomalies leading to persistent changes in the worldview. He clarified that for transformations to occur, alternative viewpoints are needed from others. Additionally, he emphasized that successful transformations might necessitate assertiveness. He has suggested pedagogies that utilize experiences via critical reflection \citep{Mez81}. \citet{Mez78} linked perspective
transformation with Freire, who emphasized the importance of becoming
reflectively aware of one's oppressive conditions to be able to take action \citep{Fre05}. %

Studies citing TLT as their framework in computing education represent several areas, including programming learning: Transformative learning was cited in a study aiming to emancipate participants for a better future by supporting digital literacy \citep{SanEtal20}. It informed a study addressing how technology can help transform discomfort in the area of HCI \citep{HalEtAl14}. Further, TLT was linked to the software engineering students' encounter with experiential-learning setting \citep{Sch07}, which can also be observed in studies not specific to computing \citep{Tho09, ForEtAl19}. It was also referred to understand whether students gained persistent learning from a project-based course (which is also about experiential learning) in the sense that this could guide them during later, more complex projects \citep{IsoNyl19}. As for programming learning, we mention ``transformative experiences'' cited in studies that investigated threshold concepts in CS \citep[e.g.,][]{MosEtAl08}. However, these studies are based on the work by \citet{MeyLan05} and address conceptual learning of the subject rather than learning at a more holistic level. Along the lines of Mezirow, \citet{LakEtAl21} used TLT as a framework to analyze students' experiences of CS0. The authors used guided discovery learning (see \citep{May04}) and extreme apprenticeship \citep{VihEtAl11}, and their interview data demonstrated mastery learning orientations in place of mere performance orientations. The authors concluded that a lot could be done already in a preliminary course to change students' orientations. %

Related work on training the TAs was extensively reviewed by \citet{Mirza2019UndergraduateTeachingAssistants}. The review reported that training had occurred both informally and structurally via courses, and contents had included learning and teaching styles, as well as communication skills, grading, professionalism, and how to encounter challenging situations.
Additionally, some studies demonstrate how mentoring complements TA-based support. \citet{MilKay02} introduced a mentoring program in which senior students volunteered to support freshmen through one-to-one relationships. The collected feedback indicated a well-received arrangement: mentors were valued as approachable and knowledgeable, and mentors perceived their role as rewarding and valuable. As raised in the introduction, we identify the study by Pon-Barry et al. \citep*{PonEtAl17} as a closely related work. These authors trained advanced students to act as near-pear mentors. Activities included written code reviews and frequent in-person feedback. Also here the mentoring was highly appreciated, e.g., regarding support for self-efficacy and learning the contents, and the mentors valued the experience. The key aspect was fostering beginners' self-regulation, acknowledging that such investment may not be easily made by faculty teachers.

\section{The study} \label{sec:method}

We position our study as follows. The research motivation with respect to including a counseling perspective in core course activities and the use of TAs was noted in the introduction already. As for SRL, the present study adopted a holistic approach through an inductive exploration of what students raised as important. As such, our approach can contribute by complementing and combining the perspectives in the literature (Section \ref{sec:theory-and-related}). As for TLT, we concluded that Mezirow's views cited in Section \ref{sec:theory-and-related} provide us with sufficient ground to reflect our results. 

\subsection{Course context}

The research was carried out at the University of Jyväskylä, a diverse university, encompassing fields such as science, mathematics, business, economics, humanities, social sciences, education, psychology, sports, health sciences, and information technology, with CS as one of its study tracks. In their first year, CS students typically enroll in CS1 and CS2 courses, algorithms, databases, and data management, as well as computer and computer networks as tools, data networks, and web publication courses. Additionally, students have the option to begin minor subject courses across various disciplines offered by the university, leading to personalized study plans. The bachelor's degree program is designed to be completed within three years.

The study was conducted in the 6-credit (ECTS) CS1 course during fall 2022. The course is open to all students at the university, but it is explicitly designed with CS majors in mind. In fall 2022, 371 students registered for the course. The course unfolded over 11 weeks and included weekly activities, a course assignment (programming a game) and an exam.
Each week began with two lectures, offered both in-person and online, where students could interact dynamically with the lecturer through quizzes and discussions on interactive examples, aiming to deepen their understanding of the weekly topics. Following the lectures, students undertook a series of 10 to 20 graded exercises as weekly exercises, which varied in difficulty from basic visual tool-assisted tasks to complex programming tasks that may require self-regulated learning. These exercises were designed to consolidate the theoretical concepts introduced in the lectures and were essential for passing. The students needed to collect at least 40\% of the total weekly exercise points, with certain exercises being marked as required to complete. Additionally, the students worked on a course assignment, typically a game or console application, throughout the entire course to demonstrate their ability to design and develop a simple application.

The course provided basic support channels present in other CS courses.
Throughout the course, the students could attend weekly lab sessions with TAs. During the lab sessions, students could work independently while being able to ask for help with the weekly exercises or the course assignment at any time. Further, the students were required to present their course assignments to TAs in one-on-one supervision sessions, allowing for iterative feedback and learning.
The students were required to attend three such sessions during the course, but the maximum number of sessions was basically unlimited. Finally, each week culminated with a review session led by the lecturer, where students could compare their submitted weekly exercises with model answers and discuss them in a collaborative environment. Overall, the support channels in the course primarily helped students with the weekly exercises and course assignments.

\subsection{Research setting}

We originally aimed to evaluate an arrangement for introducing study support into the core course activities. The arrangement included educating TAs, asking students to write reflections about advancing the course assignment and doubling the minimum number of one-to-one sessions to have enough setting for the intended research. Section \ref{subsubsec:ma} outlines educating TAs and Section \ref{subsubsec:mb} describes the student's role. Our research target was finally not to evaluate the setting in itself but through this setting develop an understanding of the students' self-regulatory processes including the use of supervision. With this respect, Section \ref{subsubsec:mc} describes the main focus of the present study.

\subsubsection{Educating TAs} \label{subsubsec:ma}

The requirement to work as a TA for the studied CS1 course is passing CS1 previously with good grades. TAs were initially selected via job interviews conducted by the course lecturer. Of the ten TAs hired to work in the course, three opted in for the arrangement in the present study. The participating TAs were also reimbruised for the additional working hours required by the arrangement. Education of the participating TAs occurred in a half-day seminar consisting of three parts. First, typical study difficulties of first-year students were presented. Second, principles of good supervision were introduced and discussed. Third, a rough script for one-to-one supervision between TAs and students was sketched to help TAs enter into dialog that includes aspects of studying. 

Regarding the first, an overview was sourced from the research conducted by the faculty on first-year students' reflections on study difficulties \citep{Isomottonen2020ExploringStudentsIdentity}. This was followed by specific attention to self-regulated learning with the help of depiction of SRL process according to \citet{Zimmerman2008InvestigatingSelfRegulationMotivation} and a recent overview of SRL strategies in computing education context following a review of \citet{Garcia2018SystematicLiteratureReviewa}.

Regarding the second, supervision was addressed through principles which Peavy's \citep*{Pea98} sociodynamic counseling approach drew from Kierkegaard's philosophy \citep[see in particular][]{Kierkegaard1859PointViewMy}. Where Kierkegaard identified himself as a religious author, Peavy acknowledged his ideas in a counseling context. The principles are as follows \citep[p. 30]{Pea98}: 
\begin{enumerate}
\item Listen, and listen from the viewpoint of the other
\item Exercise patience and humility
\item Begin where the other is
\item Let the other teach you
\item Restrain your own vanity and your need to be viewed as superior in knowledge and skill to the client
\item Be willing to admit your own ignorance
\end{enumerate}
We opined that these principles serve well in computing education as they clearly draw attention to how to meet the students who, as reported in the literature, readily demonstrate self-efficacy issues in reference to learning to program \citep[see, e.g.,][]{RogSco10,GorOro20}. The principles were translated into Finnish and slightly reworded for our purposes. 

Regarding the third, the participating TAs were given a script to help them conduct the one-to-one sessions with attention to the assignment and, importantly, the student's study habits. The session was proposed to begin with a warm-up small talk to relax the atmosphere.
Next, emphasis was suggested to be placed on addressing the assignment itself. Following this, discussions were encouraged regarding the students' study strategies and processes. Lastly, the session was intended to conclude by exploring future steps related to the assignment. It should be noted that the script was not given as a rule but as a reminder of essential foci.
The new approach to one-to-one sessions differs from the status quo of the studied CS1 course, where TAs merely review students' programming problems and evaluate the course assignment.

\subsubsection{Setting for student participants} \label{subsubsec:mb}

Participating in this research required students to attend six (as opposed to ordinary three) one-to-one supervision sessions. Additionally, participants had to log their experiences of advancing the course assignment, challenges encountered, and learning experiences in general in between the sessions, that is, during their independent studying. The students were informed that these markings are available for TAs to support the one-to-one supervision sessions. The sessions were not recorded to allow for natural interaction.

\subsection{The interview study} \label{subsubsec:mc}

The research approach can be explained in a way that SRL scoped the study but the analysis was conventional (inductive), giving attention to any emerging viewpoints. This finds support for instance in the study by \citet{San10} who noted that inductively performed analysis may emerge from the directed starting point. The analysis was based on semi-open to open interviews. Participants were given space to tell their own stories and the resulting themes were not available as interview prompts because they emerged during the study. In this connection, it is essential to note that quantification is likely not an appropriate goal \citep[cf.][Figure~1]{VaiEtAl13}.  Basing meaning on quantification in this scenario would lead to an assumption that a student would find relevant only the aspects they happened to raise or were capable of elaborating on; we cannot know how a particular student would relate to other themes indicated by other students. Another point of view is sourced from \citet{MarYar04} who noted that it can be misleading to rely on quantification with small samples in qualitative studies as such can lead to wrong assumptions of generalization on the whole population. For the above reasons, we consider quantification inappropriate in our study which is largely based on students' articulation of their personal study histories. However, \citet{MarYar04} added that researchers may provide information on which themes were rare or frequent, e.g., by using terms such as ``few'', ``most,'' etc. In the present study, we use ``one student'' only to refer to a specific example and ``some students'' when providing examples from more than one student. As is common in qualitative studies, the quotations are selected examples; how many of them are used at a particular point does not indicate count. Additionally, we followed the presentational means proposed in the method guidelines, that is, thematic maps that arrange themes qualitatively at several levels of abstraction \citep{Att01,BraCla06}.

The student participants were invited to personal interviews conducted at the end of the course. The students were informed of the research and use of personal data following the university's regulations. 
Altogether, 27 students attended the arrangement, and 14 of them opted in for the interview study.  All participants were rewarded with one extra ECTS credit for the additional effort.

The interview protocol included the following items: 
\begin{itemize} 
  \item background for participating in CS1;
  \item expectation of supervision sessions;
  \item how the student experienced the sessions;
  \item experiences with the weekly reflection task;
  \item participation activity during the supervision sessions;
  \item the influence of the sessions on study habits;
  \item and number and length of the supervision sessions. 
\end{itemize}
Of the 14 interviews, six were attended by both authors to develop a shared baseline understanding of the material, and the second author conducted the remaining eight. The interviews lasted 60 minutes on average and were transcribed. An illustrative interview excerpt is provided in Appendix \ref{app:illustration}. 

Demographic information was not devotedly collected to protect students' identities in a qualitative study, which addresses students' personal backgrounds. However, to illustrate the context of the themes, we summarize the students' background based on our first interview question as follows: CS1 was compulsory for 11 students, whose majors were CS (4), IS including cybersecurity (4), Engineering (1), and educational technology (2). All but IS have lots of similarities in the contents of bachelor studies, with IS being less technical. Of the remaining three, two students were from humanities and had selected CS1 to support their major (music technology, English teacher with technical documentation identified as a future job option). The one remaining studied social sciences but was about to change the major to cybersecurity in IT faculty, and had therefore started to take CS courses. CS1 was a compulsory course in that future target. We can summarize that three students were changing or had recently changed their field of study to CS. %
Our results conceptualize aspects influential to studying CS/programming (see Figure \ref{fig:cs1_thematic_analysis}).

The thematic analysis commenced by identifying relevant areas in the transcribed data and low-level themes. With the inductive approach, this means that the analyst is open to any perspectives considered to fit the research scope, and not looking at the data through a predefined scheme.  Consider ``Experimenting based'' in Figure \ref{fig:cs1_thematic_analysis} as an example at this level. In this step, the data were divided between the two authors, each independently analyzing seven interviews. 
Next, themes at higher levels of abstraction were identified based on observing similarities and differences among the lower-level themes. A starting point for this grouping was the suggestive connections in the data drafted by the first author during his independent analysis; we include a recreation of his hand-written drafts for the interested reader in Figure \ref{fig:suggestive-connections} in Appendix \ref{app:illustration}. The grouping work principally occurred in four half-day shared sessions, during which all independent codings were reviewed and overlappings were resolved.  The final scheme mostly includes three levels of abstraction. An example of a second-level theme is ``Learning style'' in Figure \ref{fig:cs1_thematic_analysis}, and an example of the highest level of abstraction is  ``Recognition'' in the same figure. The process in shared sessions was altogether iterative and non-linear, meaning that the data was consulted and the overall thematization (e.g., organization and labeling of themes) was revised alongside reviewing the codings. The authors worked on a whiteboard as this allowed the re-organization of the themes during the process.
We argue that the described multi-author work contributed to the trustworthiness of the study, which is elaborated on in Section \ref{subsec:trustworthiness}.

\section{Results} \label{sec:results}

\begin{figure*}
  \centering
  \makebox[\textwidth][c]{\includegraphics[width=1\textwidth]{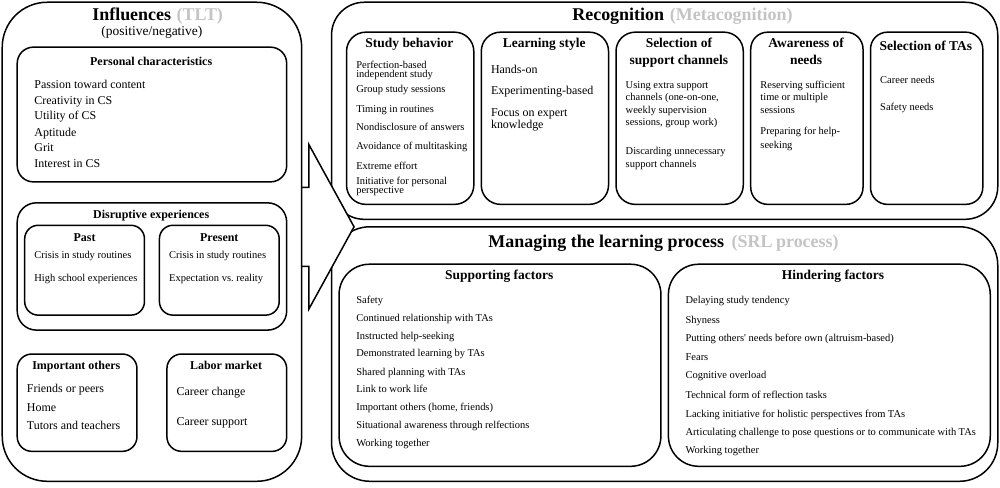}}
  \caption{\emph{Influences}, \emph{Recognition}, and \emph{Managing the learning process}. Comparison with TLT and SRL included.}.
  \label{fig:cs1_thematic_analysis}
\end{figure*}

Figures \ref{fig:cs1_thematic_analysis} and \ref{fig:supervision_needs} present the thematic analysis in three-layer maps. Figure \ref{fig:cs1_thematic_analysis} encompasses the main analysis and Figure \ref{fig:supervision_needs} adds an analytical view that differentiates between students' supervision needs. The pseudonym (userX) at the beginning of each quote refers to the participant number among those students and TAs who originally agreed to participate in the supervision arrangement. 

\subsection{Influences} \label{subsec:influences}
Students' narratives were anchored to their \emph{personal characteristics}, \emph{disruptive experiences} in the past or the present, \emph{important others}, and \emph{labor market} (see Figure \ref{fig:cs1_thematic_analysis}). These were influential factors that we, by reflection, associated with transformative learning.

\subsubsection{Personal characteristics} Passion toward the CS content was acknowledged as a starting point for studying CS. The creativity attribute, in turn, shows that the possibilities granted by programming resonate with oneself. Utility indicates that one enjoys automating varying tasks by programming. Aptitude refers to self-awareness of one's logical thinking skills concerning CS. Grit shows as insisting on persisting and not giving up, including retaking the course with more dedication. Interest indicates that interest in programming had been triggered and provided the starting point for studying. The quotes below further exemplify the personal characteristics we identified.
\begin{quote}
  [usr10:] \emph{There is a lot of freedom to express yourself in
  programming, especially when you start creating your own projects.} [illustration for creativity]
\end{quote}
\begin{quote}
  [usr28:] \emph{I like to automate things. I love building things that make life easier.} [illustration for utility] 
\end{quote}
\begin{quote}
  [usr21:] \emph{It has been decades since any school, so I don't remember anything [about maths]. [...] I remember loving maths at school. [...] I still have the aptitude for logical thinking, so I hope it will return to me [during the CS1 course.]  } [illustration for aptitude] 

\end{quote}
\begin{quote}
  [usr17:] \emph{
    This wasn't the first time I attempted to complete this course [...]
    But this time I [...] left this autumn completely empty so that I could just take my time and properly pass it.} [illustration for grit] 

\end{quote}
\begin{quote}
  [usr12:] \emph{
    I think I have always had some kind of interest [towards programming],
    but then I felt that I could maybe never learn it [...] Then I thought
    that since I'm already at the university, I could give it [CS1 course] a go
    just out of interest to see how it is. [...] I just went ahead
    and tried it.
  } [illustration for interest] 

\end{quote}

\subsubsection{Disruptive experiences} \label{subsubusec:disruptive-experiences}
Self-regulation was underpinned by \emph{past} experiences, including noticing ineffective study routines in previous courses or attempts with CS1 and, second, high school experiences. The former is illustrated below. %
\begin{quote}
  [usr23:] \emph{
    Back when I tried CS1 for the first time, I was still a bit of a lazy student. I was used to passing all my courses [...], that you just attend in the course activities a bit and get a passing grade.
    When I first dropped out, I had to do a bit of self-exploration, re-develop my learning skills.
  }
\end{quote}
\begin{quote} [usr15:] \emph{
  [The learning improved routine] evolved through frustration and wasting time. This [personal] development has come with a bitter and high cost.
}
\end{quote}
For the latter, high school experiences included references to bad teaching, making one consider if CS is something that can be learned. 
\begin{quote}
  [usr15:] \emph{
    We had computer science as a voluntary subject in secondary school.
    [...] we had an awful teacher, we just had to copy code from the blackboard.
    [The teacher] was mocking and mean, and got easily irritated. [...]
    After I finished [secondary school] I thought, ``surely I will never
    study anything related to computers.'' [...] I was suspicious [of CS1] at first.
  }
\end{quote}
Another example is a downplayed view of oneself regarding maths skills:
\begin{quote}
  [usr4:] \emph{
    Of course, programming is not directly mathematics, but they have a lot in common. You need logical reasoning.
    In high school I studied advanced math courses, but then right at the very end I dropped out. [...] The whole thing left a bad taste in my mouth.
  }
\end{quote}
The students also reported similar lived experience-based effects occurring during the \emph{present} course. These included an ineffective, stubborn routine with the exercises and learning to realize own limits with too a perfectionist approach to studying. The latter (Expectation vs. reality in the figure) is illustrated below.
\begin{quote}
  [usr25:] \emph{
    I wanted to challenge myself. Since I am such a perfectionist,
    it was a big lesson for me to accept that I would get just a minimal passing grade from this course. It is a big deal for me, really, since my
    worst grade is probably 3 [on a scale of 0--5].
  }
\end{quote}
\subsubsection{Important others} \label{subsubsec:important-others}
Students referred to important others they had observed or who had helped them (friends or peers, home) while also those who had informed and even cautioned them about the required effort (tutors and teachers in particular). Examples are provided below.
\begin{quote}
  [usr3:] \emph{
    My father works in IT, and it has always been part of my life. [...] I explained what I was interested in to my father, and he helped me [with picking university studies]. [...] [During the CS1 course] I mainly asked help from my father.
  } [illustration for home]
\end{quote}
\begin{quote}
  [usr10:] \emph{
    [...]there was a pre-course meeting with our university tutors, and there they urged us to go to all the weekly lab sessions and take all the help we need in CS1 [...] they primed us to get help from anywhere we can get it.
  } [illustration for tutors and teachers]
\end{quote}
\subsubsection{Labor market} \label{subsubsec:labor-market}
This theme encompasses narratives of career change initiated or advanced and complementing the current career by IT (CS1 required). These appeared to have a strong influence on studying CS1. An example linked with a career change is attempting to guarantee progress by doing everything possible, including learning from extracurricular materials:
\begin{quote}
  [usr24:] \emph{
    The main reason for studying CS1 was my current career change.
      [...] I was very eager to learn [CS]. Especially since I'm already in my thirties, so there is a lot at stake [regarding career change], so I wanted to really put effort into my studies. [...]
    I would have happily attended [unused session] and discussed, for instance, focusing on the advent of code, if you know it. %
  }
\end{quote}
Another example is retaking the course with increased attention, regardless of difficulties with programming, to avoid a failure that would undermine personal interest to complement the career.
\begin{quote}
  [usr17:] \emph{
    I had some mandatory courses as part of my Master's programme, and this course was one of them. [...] I think I'd benefit greatly [from the programme related to my work] in the future. [...] I don't like quitting. I was really frustrated that my previous attempts at CS1 failed because it was the only course I failed in my studies. [...]
    I decided to push through. %
  }
\end{quote}

\subsection{Recognition}
This thematic map illustrates students' recognition of their studying, encompassing \emph{Study behavior}, \emph{Learning style}, \emph{Selection of support channels}, \emph{Awareness of needs}, and \emph{Selection of TAs}.
We mainly associate this map with metacognition in SRL.

\subsubsection{Study behavior}

Recognition showed in both positive and negative study behaviors. 
Timing in study routines was raised in several ways. One had learned being able to study before noon, which had developed into a routine:
\begin{quote}
  [usr23:] \emph{
    ...I tried to do weekly tasks during two or three days always before noon,
    because I always study before noon; it's my very strong routine.
      [...] when I was writing my Bachelor's Thesis [in another field], I learned that it is impossible for me to try and study anything in the afternoon or in the evening.
  }
\end{quote}
On the other hand, studying and aiming for perfect grades at the cost of personal time was also described:
\begin{quote}
  [usr25:] \emph{
    I could not get sleep at night because I was just thinking
    [about the course and its assignments]. [...]
    When I got all the weekly exercises done, I was happy. [...]
    But it is just how I am [...] I want to do my best.
    [...] I cannot rest until I get it all done.
  }
\end{quote}
Other examples show students consciously adopting routines, such as starting weekly exercises early to come up with questions, or using group settings to collaborate without sharing answers. The latter is exemplified below:
\begin{quote}
  [usr4:] \emph{
    I was in a group of three students that are in the same situation [major and CS1 course] as me. [...] We have helped each other [during the CS1 course.] Not too much, because otherwise with friends it then is less about teaching and more about just sharing answers directly, which does not help learing.
  }
\end{quote}
Students may also avoid multitasking across courses and put in extreme effort, including extra-curricular programming, to ensure progress. Additionally, a student could consciously take the initiative during supervision sessions
to enhance the benefit of one-to-one time with the TA:
\begin{quote}
  [usr24:] \emph{
    It was nice to meet TAs [in supervision sessions]. [...] I also asked a lot besides the course assignment, but still related to programming.
    Like, ``what could I do next'' or, ``if you were me, what would you learn next in the course'' or, ``what do think my shortcomings are in this task'', etc.
  }
\end{quote}

\subsubsection{Learning style}

A conscious preference for hands-on learning instead of a passive role was emphasized. This was indicated by a student's valuation of the course assignment as a valuable activity. This meant working hands-on on the same matter for a longer time, supporting learning. Other cases indicated previously noticed awareness of this preference, such as in the following example:
\begin{quote}
  [usr24:] \emph{
    My attitude towards the course was that you must do tasks until you learn it right. [...] [This is why] I did weekly exercises so much. I think
    I learn it best when I practice the same thing over and over.
    [...] I learn by doing it [programming].
  }
\end{quote}
In case of a career change, a previous field may have had accustomed the student to hands-on learning, or students had simply observed hands-on to be more effective:
\begin{quote}
  [usr24:] \emph{
    [...] my previous studies have been hands-on. [...] [In my previous work] learning comes from work and actually doing the thing you set out to learn.
  }
\end{quote}

The experimenting-based style appeared to sync with creativity and an acknowledged preference for experimenting instead of following the course materials, as illustrated below.
\begin{quote}
  [usr23:] \emph{that my friend can do practical things [...] just during free-time [...] it was like, this [programming] could be quite interesting. I didn't want to [study programming] with deliberation or with deep thought.
  I just test things that seem interesting to me. If it works, great.
  If it doesn't, then I look at how to fix it.
    [...] I didn't attend any lectures. [...] %
  }
\end{quote}
Yet another conscious style was attempting to focus on learning and the expert knowledge that could be gained from course activities.

\subsubsection{Selection of support channels} \label{subsubsec:selection-support-ch}

A clear indication of recognition was that the study participants were aware of
potential self-regulation challenges and hence selected and appreciated the course activities. A prominent example is the extra one-on-one supervisions sessions which themsevles were part of the study (see Section \ref{subsubsec:mb}). One student was even amused after being told that only a small portion of the course students participated in the study:
\begin{quote}
  [usr23:] \emph{
    I remember I was amused a lot [...] as to why more people did not use one-to-one support. To me, it looked like you get free remedial lessons.
    It's like you get a free VIP pass to get personal help for your studying.
  }
\end{quote}
Conversely, another kind of scenario was that the students consciously skipped the voluntary activities they did not experience a need for. 
\begin{quote}
  [usr28:] \emph{
    I did almost all course assignments by myself. I attended lectures only during the first period. In the second period, I just did the weekly exercises. [...] I went just to one weekly lab to do the mandatory debugging exam.
  }
\end{quote}

\subsubsection{Awareness of needs}
This aspect conveys that students are aware of their needs, which influences their action-taking. On a larger scale, some students devoted their whole fall semester to CS1, leaving out other courses.
\begin{quote}
  [usr23:] \emph{
    I knew this course was really challenging [...] so I purposefully had this course as the only one during the whole fall semester. I thought this was such an important course to me that I didn't want to risk dropping out.
  }
\end{quote}
\begin{quote}
 [usr21:] \emph{
   I reserved plenty of time [for the course] [...]
   I had just two concurrent courses at the start of the fall and a bit more
   at the end. [...] I used up all the time I had reserved for the course.
 }
\end{quote}
As for supervision, students could consciously reserve multiple one-to-one sessions back to back to guarantee help. Another scenario was that students arrived well-prepared at one-to-one sessions based on the awareness of their current needs.

\subsubsection{Selection of TAs} \label{subsubsec:selection-of-tas}
The recognition showed in how students chose TAs. Advanced students noted that they started returning to more experienced TAs with whom career needs could be addressed: 
\begin{quote}
  [usr24:] \emph{
    In the end, \underline{I began to talk to those TAs} from whom I got the most help.
    After all, it's a bit pointless to ask about a problem -- like how to construct objects -- if the TA doesn't understand the concept.
    [...] My supervision session felt like I was at a real workplace. You have a problem; you think about it, then leave it for the time being. Then you consult a colleague who gives either code snippets or suggestions.
  } (Emphasis added.)
\end{quote}
\begin{quote}
  [usr28:] \emph{
    I was a self-taught programmer showing my course assignment [to TAs].
    I wanted to get feedback on whether I had any bad coding patterns.
    [...] We talked about how things are done in different languages.
  }
\end{quote}

On the other hand, experiencing safety made inexperienced students continue with TAs with whom they had experienced it. 
\begin{quote}
[usr21:] \emph{
  \underline{That's why I chose [TA's name] right away}, because it felt like we clicked during the weekly labs. The communication has been really natural, so in my opinion, it has also helped that I've dared to ask, dared to say that I really don't understand, like ``can you please explain it in some other way so that I understand.''
} (Emphasis added.)
\end{quote}

The theme is explored further when conceptualizing differing needs in Section \ref{subsec:differing-needs} and Figure \ref{fig:supervision_needs}.

\subsection{Managing the learning process}
This thematic map presents \emph{Supporting factors} and  \emph{Hindering factors}. We primarily associate this map with the SRL process -- that is, the planning, action-taking, monitoring, and reflection processes.

\subsubsection{Supporting factors} \label{subsubsec:supporting-factors}

It was easy to read that safety played a key role for students and supported their progress by lowering a participation threshold. The continued relationship with particular TAs, mentioned in the previous section, was also stressed as a supportive element and positive feedback on the whole research arrangement. The example below illustrates both of these aspects.
\begin{quote}
  [usr17:] \emph{
    I had [TA's name] in my one-to-one sessions.
    They provided a safe environment. [...] They helped me push forward and told me what to focus on next. It was just what I needed.
      [...] I liked them, and it was easy to talk to them.
    I think it's easier to continue [the supervision sessions] with someone who knows about my situation, who even remembers me.
  }
\end{quote}

An interesting aspect was that help-seeking became easier as an instructed task, e.g., TA advising the student to the weekly lab session to ask particular advice. It seemed to help to overcome fears.
\begin{quote}
  [usr3:] \emph{
    Once, one of the TAs said that he was in the weekly lab sessions, and we agreed that I would go there.
      [...] A lot of times in the one-to-one sessions, I was encouraged
    to go to the weekly labs since I could get help there too. [...] It was good when someone said, ``go there and ask this''.
  }
\end{quote}
Students also mention the value of observing how TAs searched for information, encouraging students to do the same independently.
\begin{quote}
  [usr15:] \emph{
    The TA showed me [an example] by googling something some webpage [...]
    I later tried to search the same thing but used a different search result, because I didn't want to copy [their approach] directly and learn myself.
  }
\end{quote}
Shared planning during the one-to-one sessions was valued a lot. It appeared as if TAs helped the students quite directly to regulate, while it also seemed that beginners were cognitively overwhelmed and simply did not know the sensible next step in the assignment.
\begin{quote}
  [usr3:] \emph{
    It was nice [in one-to-one sessions] that at the end we always agreed on the goals for the next session. [...] It helped me a lot to set these ``in-between deadlines''.
  }
\end{quote}
\begin{quote}
  [usr10:] \emph{
    [...] often I felt I didn't know how to proceed, how to subdivide the work [...] So after every one-to-one session we planned with the TA what I should do next, like ``tell me what to do next, and I'll do it''.
  }
\end{quote}
The link to working life helped advanced students find supervision meaningful, as they received prospects for how course content was linked with software engineering.
\begin{quote}
  [usr24:] \emph{
    [I preferred] especially supervisions, where the TA had more expertise from the work life and knew how the things taught in CS1 can be related to
    project management or other concrete work.
  }
\end{quote}
Important others were mentioned as a concrete support resource: the help was received from home and friends. The help from home was shown in an earlier example (Section \ref{subsubsec:important-others}), while the below examplifies friends. 
\begin{quote}
  [usr1:] \emph{
    ...I didn't feel like I needed a lot of TA support.
    But this is because I had friends who also studied this course with me.
      [...] it's easier to open up to my friends.
  }
\end{quote}
\begin{quote}
  [usr25:] \emph{
    I had a friend who studied the course before, so I could ask them quickly
    via phone if there was some problem I could not get past.
  }
\end{quote}
Situation awareness emerged from asking students to track their progress into the virtual learning environment, a task included in this research.  Students noticed the value of the awareness obtained, with some students noting that they had learned to appreciate the conduct before.
\begin{quote}
  [usr12:]  \emph{
    [...] it [writing reflections] kind of helped me understand my progress at each point of the course.
    It was also nice to see how much I improved at the end of the course.
  }
\end{quote}
\begin{quote}
  [usr10:] \emph{
    ...I was really used to self-reflection and writing down my thoughts and observations. [...] If I solved some problem, then I also wrote down about my successes. It worked as a reminder, like ``hey, you struggled a lot with it but still got it done''.
  }
\end{quote}
Working together in sessions or informally was supportive. It could provide psychological support or shared work on exercises.
\begin{quote}
  [usr12:] \emph{
    I was in a study group and participated in one-to-one supervision sessions. They both helped me understand weekly tasks and get them done.
      [...]
    It helped to think aloud with someone else; we could combine our knowledge and different perspectives [...]
  }
\end{quote}

\subsubsection{Hindering factors}

Students raised their tendency to delay their work, e.g., when reasoning why they selected to participate in the research with additional support. 
\begin{quote}
  [usr10:] \emph{
    Yes, those [one-on-one supervision sessions] paced the [course assignment] in a completely different way than if there had been just the three mandatory submissions. Otherwise, it would have gone even more in the way that I'm there on the last night, struggling before the deadline. [...]  Project-based independent work assignments are not exactly my strength yet.
  }
\end{quote}

Another kind of hindering factor was shyness influencing help-seeking: 
\begin{quote}
  [usr3:] \emph{
    [In weekly lab sessions] there were a lot of students present.
    I saw that TAs were very busy anyway.
    I thought that maybe I'll try to do tasks by myself. [...]
    I'm a bit on a shy side, so it was difficult to ask for help especially
    if I thought that I should already know this [topic].
  }
\end{quote}

An interesting aspect was also the altruistic behavior. With altruistic behavior, a student did not want to take time from others and skipped their needs in weekly lab sessions, and could avoid going overtime in one-to-one sessions regardless of their needs. Both aspects are illustrated below.

\begin{quote}
  [usr17:] \emph{
    The last time I was in weekly labs, I worked with someone as a pair.
    But then I felt uneasy because my starting level was much lower than my peers'. I felt I just slowed them down.
  }
\end{quote}
\begin{quote}
  [usr2:] \emph{
    I was very precise with time [in one-to-one supervision sessions]
    because I was aware there were others after me [...] so I could say [to TA] that I didn't have any more questions even though I did [...] so we could stay within the time frame.
  }
\end{quote}

Fears were noted relating to doubts about oneself, which can affect participating in course activities: 
\begin{quote}
  [usr2:] \emph{
    [...] of course you should ask help if you don't know something.
      [...] I had an impression at the beginning that everyone else already knows a ton about CS, so... you begin to feel ashamed because you think the TA will just say ``oh, you don't know that'' or ``well, it may be pretty difficult for you''...
  }
\end{quote}

Relatedly, the cognitive load was present in the student becoming overwhelmed by course contents and therefore losing the ability to seek help. \begin{quote}
  [usr2:] \emph{
    I didn't always know what I should be knowing after each week [...]
    I tried to do weekly tasks and thought if I should go to weekly lab sessions, but then I started thinking, ``okay, I don't understand anything what [the lecturer] talked about and I didn't find any mentions in the materials; am I really that stupid to not understand anything?''
    So I just thought that I did't dare to seek help, I'd rather sit home and bang my head against the wall until I find the solution.
      [...] When you're under time pressure, you can't rationally approach the situation; you just see that you can't do it.
  }
\end{quote}
Other cases we identified relating to cognitive load referred to the challenge with programming thinking. In our interpretation, the shyness, altruism, fears, and cognitive overload indicate a kind of self-handicapping in help-seeking.

Another kind of challenge was the form of the reflection task. Students noted a preference for pen and paper or free-form tracking and were reluctant to use the guided format in the virtual learning environment.
\begin{quote}
  [usr15:] \emph{
    I personally don't like these computer [reflections], so I didn't write them there. I just write with pen and paper. I have [all the reflections] in my notebook [...]
  }
\end{quote}
It could be also extracted that supervision sessions lacked TAs' initiative for holistic dialog taking into account students' background and current personal needs outside the assignments.%
\begin{quote}
  [usr15:] \emph{
    Well, [the TAs] didn't really talk [about learning experiences]; I just
    started talking about my experiences myself. Of course, some of them related to me, but I didn't get any questions about ``my learning''.
  }
\end{quote}
Overlapping with a cognitive load perspective, the challenge was articulating questions in help-seeking with a yet incomplete understanding. A similar effect was raised in the correspondence between a student and TA: a lack of understanding of what was discussed.
\begin{quote}
  [usr4:] \emph{
    [...] the TA might be really skillful [in CS], but they could not word the concepts in a way I would understand them. The same goes the other way.
    These situations were ones where I could not make progress.
    I was frustrated; was I really that bad? Or did I not just understand them?
  }
\end{quote}
Finally, working together was also noted as a difficulty because it was considered disturbing compared to working alone, or was experienced to take time without receiving help for oneself.
\begin{quote}
[usr23:] \emph{And for once, I asked for help, but then the one who came was another student, and after that, my whole rest of the day went to that, well not the whole rest of the day. [...] I started to help that other student because I had completed the task where [the other student] needed help. And then I felt a bit frustrated because my study time had gone to that, so my own goal in a way, my own problems hadn't progressed at all. But at the same time, I was really tired because I had been advising that other person.}
\end{quote}

\subsection{Differing superivision needs} \label{subsec:differing-needs}

\begin{figure}
  \centering
  \includegraphics[width=0.7\textwidth]{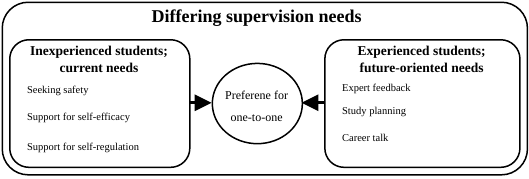}
  \caption{The analytic view of differing supervision needs.}
  \label{fig:supervision_needs}
\end{figure}

By comparing the cases, the analysis suggested differing needs in supervision (see Figure \ref{fig:supervision_needs}). The inexperienced students struggling with programming learning were relieved about the safety and the received understanding of their progress (cf. self-efficacy support). Additionally, their needs emphasized the help for planning the work with assignments, given the cognitive load they were encountering (self-regulation support). The inexperienced students' needs appeared to relate to their current course situation. On the other hand, those with experience (e.g., strong hobbyism or effort due to a labor market-related incentive) emphasized the need for receiving expert feedback comparable to a colloquial professional dialog, attention to planning further studies, and information about working life expectations (career talk). These needs focus beyond the course toward the future. In both scenarios, one-to-one supervision was valued over lab sessions: the inexperienced found the one-to-one sessions safer, whereas the experienced valued them as more personal while also acknowledging the safe atmosphere. The items in the figure were mostly illustrated in the previous section. A couple of more illustrations are given below.

\begin{quote}
  [usr4 regarding safety:] \emph{
    I sought safety. It [CS] was so outside my comfort zone, especially
    at the start of the course. [The one-to-one sessions] made me feel that I am not alone. [...] I sought safety for completing the course; that I can get personal help if I need especially since so many other students attend the course.
  }
\end{quote}

\begin{quote}
  [usr17 regarding support for self-regulation:] \emph{
    The TA helped me continue with the course a lot. [...] Even though I can self-regulate well, here [in CS] I couldn't because I simply did not know, what to do next. [...] I needed someone to just say, ``now, focus on loops and then you do this'' or ``go to weekly labs if you need more help''
  }
\end{quote}

\section{Discussion} \label{sec:discussion}

With the inductive data analytical approach, the answer to the research question
(\emph{How self-regulatory processes and use of supervision emerged during
CS1?}) is the resultant thematization.  Overall, students' SRL was interpreted
from their characterization of their recognition of their study habits (linked
to metacognition) and the factors that supported or hindered the learning
process (linked to SRL process). The learning approaches described by the
students largely align with the CS1 learning strategies described in prior
studies (e.g., ``timing in routines'' vs. ``goal-setting and planning'' strategy
in \citep{Garcia2018SystematicLiteratureReviewa}; ``situational awareness
through reflections'' vs. ``self-assessment'' strategy in
\citep{Falkner2014IdentifyingComputerScience}), which provides initial
validation. Further, our analysis highlights the effect of student's personal
background on their learning and SRL behavior (linked to TLT). Additional SRL
and TLT aspects of the data and the practical implications are discussed below.
With the present research approach, this discussion contextualizes the
themes by incorporating literature, while not attempting to verify theories. 

From the interviews, we observed that while all CS1 students were aware of the
need to self-regulate, some students appeared to lack the skills or even
willingness to do so.  It is known that beginner CS students tend to
self-regulate less than more experienced students, as reported by
\citet{Loksa2016RoleSelfRegulationProgramming}.  However, in the analyzed
interviews, some students were aware of their SRL challenges and explicitly
elected to have the TAs handle the SRL process for them during one-to-one
sessions (see, e.g., Section \ref{subsec:differing-needs}, usr17).
Therefore, we characterize such behavior as a sort of \emph{externalized
self-regulation}.  To our best knowledge, such behavior has not been directly
reported in computing education research, although we believe the phenomenon can be familiar to
educators in any supervision setting.  On the one hand, this behavior may be
close to a kind of scaffolding.  For example, \citet{Aze08Why} has reported
success in using instructors to help students self-regulate via external
prompts. On the other hand, our observed behavior could lend itself to social
loafing, whereby a student may reduce their study efforts in a social setting
\citep{Kar93Social}.  As such, this behavior could also be characterized as
``self-regulation free-riding'', in which students may lessen their SRL efforts
when a TA is available to support SRL.

It must be noted that our analysis revealed both the scaffolding and free-riding
types of externalized self-regulation. For scaffolding, a student's previously
unsuccessful self-regulation attempts may lead them to more actively use TAs to
ensure the correct self-regulation strategies are used (see, e.g., Section
\ref{subsubsec:selection-support-ch}, usr23). Further, being overwhelmed by the
cognitive load present in the course assignment together with a lack of past SRL
skills in CS may even force a student to ask TAs to set goals or subdivide the
assignment into smaller tasks (see, e.g., Section
\ref{subsubsec:supporting-factors}, usr10 in \emph{Shared planning}). As for
free-riding, we did not observe any direct articulations of such, but it could
be read that some students willingly externalized their self-regulation (see,
e.g., Section \ref{subsec:differing-needs}, usr17; Section
\ref{subsubsec:supporting-factors}, usr3 in \emph{help-seeking}).  Therefore,
the results suggest that while frequent shared planning with TAs can be critical
for the progress of a beginner CS1 student, they also caution against taking all
the SRL efforts from the student.  Overall, we posit that TAs should be made
aware to provide enough SRL support for beginners while encouraging students to
develop their own self-regulation skills.

It was interesting to notice that students' personal histories influenced SRL
skills and, in our interpretation, even more so than the arrangement provided.
The plausibility of this connection is supported by \citet{MorEtAl12}, who
discussed SRL in conjunction with TLT and transformative pedagogy in nursing,
and whose example narratives appear similar to ours. In short, through critical
self-reflection, their students had better taken ownership of learning. We read
from their examples that this was anchored to a personal need to progress to the
next level of education or the outcomes of critical reflection prompted by
educators. In our work, disruptive experiences (see, e.g., usr23 and usr15 in
Section \ref{subsubusec:disruptive-experiences}) are easy to read as influential
``anomalies'' referred to in TLT, and we also found career change narratives
that were not only about reacting to societal changes but acting according to
personal imperatives (e.g., usr17 in Section \ref{subsubsec:labor-market}).
Regarding the disruptions, it was interesting to see that previous difficulties
and even drop-outs from CS1 had strengthened students' SRL and not led to
discontinued studies (several students in our data). We also placed
personal characteristics and important others under TLT (see Figure
\ref{fig:cs1_thematic_analysis}), which is because TLT proposes that
perspectives should be available from others \cite{Mez78}. We accordingly saw in
the data that, for instance, the ``Passion toward the content'' was referred to
in conjunction with finding and inspecting a relative's thesis that included
some form of program code as a youngster (usr28), or that the important others
had had an actual effect on how students related to supervision; e.g., usr10 cited in Section \ref{subsubsec:important-others} continued that the suggestions by tutors had invoked attention to making sufficient effort during the course. 

Based on these reflections, we believe that the present results can inform the education of TAs, emphasizing that their role should include prompting students to engage in critical self-reflection regarding their history and goals for the course. We undoubtedly want to help students not to drop out and wait for self-awareness and improved self-regulation the hard way. This suggestion aligns with the study by \citet{LakEtAl21}, which demonstrated that certain pedagogies helped transform students' position to their learning during CS0. The other reference study \citep{PonEtAl17} reported positive feedback from using near-peer mentors, thereby supporting the use of student supervisors, but it did not at least explicitly raise transformative pedagogy as a mentor's tool.

The results arguably guide designing CS1 supervision. Firstly, the results can
inform supervision processes.  Given the influences (see Section
\ref{subsec:influences}), one-to-one supervision could start from a dedicated
discussion on the student's background and then develop into continued
relationships (see Section \ref{subsubsec:supporting-factors}). The initial
screening of students' backgrounds could be used for allocating students to TAs
according to students' needs (see Section \ref{subsec:differing-needs}). This concurrently means recruiting TAs, ensuring that some have more experience, as this would aid in allocation and potentially encourage SRL across differing student needs. Secondly, we believe that the thematic maps that were produced through the present study help educate TAs. Compared with our current effort, which informed TAs rather generally, the results encourage TAs to become sensitive to influential background factors and various supporting factors and obstacles with SRL. Thirdly, we conclude that undertaking supervision conforming to Peavy's and Kierkegaard's ideas requires additional resources and that we were initially too optimistic about the success of the arrangement. The current results indicated that a holistic counseling perspective was initiated by students and not too much by TAs. However, we acknowledge that addressing students' situations holistically was limited by the time available for the supervision sessions. Part of the data
indicated that holistic discussions were possible by going overtime or reserving
multiple sessions. %

\subsection{Trustworthiness and limitations}  \label{subsec:trustworthiness}

We discuss trustworthiness using concepts of credibility, dependability, and transferability known from \citet{LinCub85}. 

Credibility refers to the question of whether our results resonate with the participants -- that is, if they are correct \citep[p. 296]{LinCub85}. This can be advanced by building trust and the researchers' prolonged relationship with the setting \citep[p. 301]{LinCub85}. As for trust, the first author is in charge of a first-year induction course and is known by first-year students through that connection. In addition to formally informing students according to the university's guidelines, the second author introduced himself and this research in the classroom, carefully explaining our interest in hearing the students and this research as a vehicle to understand and develop supervision in the faculty. We were also well aware of our CS1 context in a prolonged manner: both authors had participated in CS1 research, and the second author had previously worked as a TA. In this particular CS1 course, the present research did not, however, mean that we are working with the study participants in a prolonged manner. 

We did not find a need to arrange peer debriefing outside the research group, which was also suggested by \citet{LinCub85}, because the independent codings were exposed and discussed one by one in the shared research meetings between the authors. We did not identify biases during these sessions. The threat of directing the analysis by theory and not proceeding inductively as intended was discussed in advance. \citet{LinCub85} also pointed to referential adequacy, e.g., by archiving data publicly. However, such approach does not suit our need to protect identities in the study where students could, for instance, elaborate on personal matters such as previous dropouts. Instead, short quotations were frequently used in the results narrative, which allows the reader to consider the fit between our thematization and data.

Dependability refers to how results depend on one or more perspectives (cf. reliability) and can be addressed by introducing multiple researchers \citep{Gub81, LinCub85} to avoid perspectivism \citep{CorEtAl14}. As depicted in the method section, both authors attended the interviews and
performed analyses independently. Additionally, the interpretations were replicated through shared go-throughs of all codings. We believe that our procedure responds to this aspect of trustworthiness. 

Generalizability in the qualitative domain can be discussed in terms of transferability, which means how results potentially apply to other settings. According to \citet{LinCub85}, this can be advanced by depicting the research setting. The potential users of the research may, then, consider the applicability of the results to their settings. We have described our setting in Section \ref{sec:method}, including general characteristics of the university, the course, and the supervision setting underpinning the research. This view of transferability hence stresses the user and not the original researchers. Our epistemological view is less relativist, and we would add, comparable to the idea of substantive grounded theories developed within particular domains \citep{GlaStr67}, that the results are likely to be relevant under similar conditions. For instance, we envisage that settings in which first core courses challenge students similar to programming courses might demonstrate similar qualitative outcomes. We also draw the reader's attention to our CS1 population consisting of both IT faculty students (CS, IS, Engineering, Educational technology) and students studying CS as their minor or taking a few CS courses to support their degree. Other universities may have CS1 courses dedicated to majors vs minors. 

The study is not without limitations. The number of interviews was small given the large size of the class. While we acknowledge this as a chief limitation, we respond to it by noting that we attempted to understand students' perspectives of their study processes in relation to available support, and not focus on impact or efficiency. The latter goal would, of course, require a larger sample size. These two modes of evaluation are acknowledged in the literature \citep{MacEtAl00}. Additionally, we observed that the research was attended by students with varying backgrounds. Thereby we opine that the thematization can nevertheless be of interest in the field.

\section{Conclusions and future directions}

The results and the above discussion provoke conclusions as follows. First, attention should be given to interactions in supervision regarding how to support self-regulation while being able to help with cognitively challenging assignments. Second, it seems essential that educators pay attention to students' personal histories to understand their SRL processes and support needs. Here, the results and the cited literature call attention to transformative learning pedagogies; in the context of the present study, this means that TAs could provoke critical reflection.  Third, we believe that the results, overall, inform how to organize one-to-one supervision processes starting from TA recruitment and can work as learning material for educating TAs.

We envisage that this research area would benefit from discourse analyses of one-to-one supervision sessions to understand the effects of interactions between TAs and students in detail. Creating and studying a safe atmosphere in lab sessions appears crucial because, currently, our results hint that safety is rather experienced in a one-to-one setting. We relatedly recognize the comfort level as a success factor in multiple studies on introductory programming (e.g., \cite{WilSch01, Ven05}) and here emphasize the need for closer analyses of interactions in sessions attended by multiple students. Overall, we identify a need for studies on holistic pedagogies that consider students' personal histories. One example is the implementation of critical CS, where students' backgrounds were taken into account, and content was prepared to present justice-themed assignments \citep{MalRes19}. This ``alternative'' pedagogy, however, led to the stigmatization of participating students. While we agree with the goals of \citet{MalRes19}, based on our results, we add thIsomat pedagogies acknowledging students' background experiences should be pursued to facilitate purposeful supervision and promote transformations towards independent self-regulation.

\bibliographystyle{ACM-Reference-Format}
\bibliography{refs_final}


\begin{thebibliography}{66}


\ifx \showCODEN    \undefined \def \showCODEN     #1{\unskip}     \fi
\ifx \showISBNx    \undefined \def \showISBNx     #1{\unskip}     \fi
\ifx \showISBNxiii \undefined \def \showISBNxiii  #1{\unskip}     \fi
\ifx \showISSN     \undefined \def \showISSN      #1{\unskip}     \fi
\ifx \showLCCN     \undefined \def \showLCCN      #1{\unskip}     \fi
\ifx \shownote     \undefined \def \shownote      #1{#1}          \fi
\ifx \showarticletitle \undefined \def \showarticletitle #1{#1}   \fi
\ifx \showURL      \undefined \def \showURL       {\relax}        \fi
\providecommand\bibfield[2]{#2}
\providecommand\bibinfo[2]{#2}
\providecommand\natexlab[1]{#1}
\providecommand\showeprint[2][]{arXiv:#2}

\bibitem[{Attride-Stirling}(2001)]%
        {Att01}
\bibfield{author}{\bibinfo{person}{Jennifer {Attride-Stirling}}.}
  \bibinfo{year}{2001}\natexlab{}.
\newblock \showarticletitle{Thematic Networks: {{An}} Analytic Tool for
  Qualitative Research}.
\newblock \bibinfo{journal}{\emph{Qualitative Research}} \bibinfo{volume}{1},
  \bibinfo{number}{3} (\bibinfo{year}{2001}), \bibinfo{pages}{385--405}.
\newblock
\href{https://doi.org/10.1177/146879410100100307}{doi:\nolinkurl{10.1177/146879410100100307}}


\bibitem[Azevedo et~al\mbox{.}(2008)]%
        {Aze08Why}
\bibfield{author}{\bibinfo{person}{Roger Azevedo}, \bibinfo{person}{Daniel~C.
  Moos}, \bibinfo{person}{Jeffrey~A. Greene}, \bibinfo{person}{Fielding~I.
  Winters}, {and} \bibinfo{person}{Jennifer~G. Cromley}.}
  \bibinfo{year}{2008}\natexlab{}.
\newblock \showarticletitle{Why Is Externally-Facilitated Regulated Learning
  More Effective than Self-Regulated Learning with Hypermedia?}
\newblock \bibinfo{journal}{\emph{Educational Technology Research and
  Development}} \bibinfo{volume}{56}, \bibinfo{number}{1} (\bibinfo{date}{Feb.}
  \bibinfo{year}{2008}), \bibinfo{pages}{45--72}.
\newblock
\showISSN{1042-1629, 1556-6501}
\href{https://doi.org/10.1007/s11423-007-9067-0}{doi:\nolinkurl{10.1007/s11423-007-9067-0}}


\bibitem[Boyd(1991)]%
        {Boy91}
\bibfield{author}{\bibinfo{person}{Robert~D Boyd}.}
  \bibinfo{year}{1991}\natexlab{}.
\newblock \bibinfo{booktitle}{\emph{Personal Transformations in Small Groups:
  {{A Jungian}} Perspective}}.
\newblock \bibinfo{publisher}{Routledge}.
\newblock
\href{https://doi.org/10.4324/9780203359099}{doi:\nolinkurl{10.4324/9780203359099}}


\bibitem[Braun and Clarke(2006)]%
        {BraCla06}
\bibfield{author}{\bibinfo{person}{Virginia Braun} {and}
  \bibinfo{person}{Victoria Clarke}.} \bibinfo{year}{2006}\natexlab{}.
\newblock \showarticletitle{Using Thematic Analysis in Psychology}.
\newblock \bibinfo{journal}{\emph{Qualitative research in psychology}}
  \bibinfo{volume}{3}, \bibinfo{number}{2} (\bibinfo{year}{2006}),
  \bibinfo{pages}{77--101}.
\newblock
\href{https://doi.org/10.1191/1478088706qp063oa}{doi:\nolinkurl{10.1191/1478088706qp063oa}}


\bibitem[Campbell et~al\mbox{.}(2019)]%
        {CamEtAl19}
\bibfield{author}{\bibinfo{person}{Jennifer Campbell}, \bibinfo{person}{Andrew
  Petersen}, {and} \bibinfo{person}{Jacqueline Smith}.}
  \bibinfo{year}{2019}\natexlab{}.
\newblock \showarticletitle{Self-Paced Mastery Learning Cs1}. In
  \bibinfo{booktitle}{\emph{Proceedings of the 50th Acm Technical Symposium on
  Computer Science Education}}. \bibinfo{publisher}{ACM}, \bibinfo{address}{New
  York, NY}, \bibinfo{pages}{955--961}.
\newblock
\href{https://doi.org/10.1145/3287324.3287481}{doi:\nolinkurl{10.1145/3287324.3287481}}


\bibitem[Cornish et~al\mbox{.}(2014)]%
        {CorEtAl14}
\bibfield{author}{\bibinfo{person}{F. Cornish}, \bibinfo{person}{Gillespie A.},
  {and} \bibinfo{person}{Zittoun T.}} \bibinfo{year}{2014}\natexlab{}.
\newblock \showarticletitle{Collaborative Analysis of Qualitative Data}.
\newblock In \bibinfo{booktitle}{\emph{The {{SAGE}} Handbook of Qualitative
  Data Analysis}}. \bibinfo{publisher}{Sage Publications},
  \bibinfo{pages}{79--93}.
\newblock
\href{https://doi.org/10.4135/9781446282243}{doi:\nolinkurl{10.4135/9781446282243}}


\bibitem[Daloz(1986)]%
        {Dal86}
\bibfield{author}{\bibinfo{person}{Laurent~A. Daloz}.}
  \bibinfo{year}{1986}\natexlab{}.
\newblock \bibinfo{booktitle}{\emph{Effective {{Teaching}} and {{Mentoring}}:
  {{Realizing}} the {{Transformational Power}} of {{Adult Learning
  Experiences}}} (\bibinfo{edition}{1} ed.)}.
\newblock \bibinfo{publisher}{Jossey-Bass}.
\newblock
\showISBNx{978-1-55542-001-7}


\bibitem[Dirkx(1998)]%
        {Dir98}
\bibfield{author}{\bibinfo{person}{John~M Dirkx}.}
  \bibinfo{year}{1998}\natexlab{}.
\newblock \showarticletitle{Transformative Learning Theory in the Practice of
  Adult Education: {{An}} Overview}.
\newblock \bibinfo{journal}{\emph{PAACE journal of lifelong learning}}
  \bibinfo{volume}{7} (\bibinfo{year}{1998}), \bibinfo{pages}{1--14}.
\newblock


\bibitem[Doebling and Kazerouni(2021)]%
        {Doebling2021PatternsAcademicHelpSeeking}
\bibfield{author}{\bibinfo{person}{Augie Doebling} {and}
  \bibinfo{person}{Ayaan~M. Kazerouni}.} \bibinfo{year}{2021}\natexlab{}.
\newblock \showarticletitle{Patterns of {{Academic Help-Seeking}} in
  {{Undergraduate Computing Students}}}.
\newblock \bibinfo{journal}{\emph{ACM International Conference Proceeding
  Series}} (\bibinfo{date}{Nov.} \bibinfo{year}{2021}).
\newblock
\href{https://doi.org/10.1145/3488042.3488052}{doi:\nolinkurl{10.1145/3488042.3488052}}


\bibitem[Falkner et~al\mbox{.}(2014)]%
        {Falkner2014IdentifyingComputerScience}
\bibfield{author}{\bibinfo{person}{Katrina Falkner}, \bibinfo{person}{Rebecca
  Vivian}, {and} \bibinfo{person}{Nickolas J~G Falkner}.}
  \bibinfo{year}{2014}\natexlab{}.
\newblock \showarticletitle{Identifying {{Computer Science Self-Regulated
  Learning Strategies}}}.
\newblock  (\bibinfo{year}{2014}).
\newblock
\href{https://doi.org/10.1145/2591708.2591715}{doi:\nolinkurl{10.1145/2591708.2591715}}


\bibitem[Ferreira et~al\mbox{.}(2024)]%
        {FerEtAl23}
\bibfield{author}{\bibinfo{person}{Deller~James Ferreira},
  \bibinfo{person}{Dirson~Santos Campos}, {and}
  \bibinfo{person}{Anderson~Cavalcante Gon{\c c}alves}.}
  \bibinfo{year}{2024}\natexlab{}.
\newblock \showarticletitle{Regulatory Strategies for Novice Programming
  Students}. In \bibinfo{booktitle}{\emph{Computer Supported Education}},
  \bibfield{editor}{\bibinfo{person}{Bruce~M. McLaren}, \bibinfo{person}{James
  Uhomoibhi}, \bibinfo{person}{Jelena Jovanovic}, {and}
  \bibinfo{person}{Irene-Angelica Chounta}} (Eds.).
  \bibinfo{publisher}{Springer Nature Switzerland}, \bibinfo{address}{Cham},
  \bibinfo{pages}{136--159}.
\newblock
\href{https://doi.org/10.1007/978-3-031-53656-4\_7}{doi:\nolinkurl{10.1007/978-3-031-53656-4\_7}}


\bibitem[Fortune et~al\mbox{.}(2019)]%
        {ForEtAl19}
\bibfield{author}{\bibinfo{person}{Tracy Fortune}, \bibinfo{person}{Shinead
  Borkovic}, \bibinfo{person}{Anoo Bhopti}, \bibinfo{person}{Renee Somoza},
  \bibinfo{person}{Ha~Chan Nhan}, {and} \bibinfo{person}{Shabnam Rangwala}.}
  \bibinfo{year}{2019}\natexlab{}.
\newblock \showarticletitle{Transformative Learning through International
  Project-Based Learning in the Global South: {{Applying}} a
  Students-as-Partners Lens to a ``High-Impact'' Capstone}.
\newblock \bibinfo{journal}{\emph{Journal of Studies in International
  Education}} \bibinfo{volume}{23}, \bibinfo{number}{1} (\bibinfo{year}{2019}),
  \bibinfo{pages}{49--65}.
\newblock
\href{https://doi.org/10.1177/1028315318814571}{doi:\nolinkurl{10.1177/1028315318814571}}


\bibitem[Freire et~al\mbox{.}(2005)]%
        {Fre05}
\bibfield{author}{\bibinfo{person}{Paulo Freire}, \bibinfo{person}{Joel
  Kurtti}, \bibinfo{person}{Tuukka Tomperi}, {and} \bibinfo{person}{Juha
  Suoranta}.} \bibinfo{year}{2005}\natexlab{}.
\newblock \bibinfo{booktitle}{\emph{Sorrettujen Pedagogiikka}}.
\newblock \bibinfo{publisher}{Vastapaino}, \bibinfo{address}{Tampere}.
\newblock
\showISBNx{951-768-159-3}


\bibitem[Garcia et~al\mbox{.}(2018)]%
        {Garcia2018SystematicLiteratureReviewa}
\bibfield{author}{\bibinfo{person}{Rita Garcia}, \bibinfo{person}{Katrina
  Falkner}, {and} \bibinfo{person}{Rebecca Vivian}.}
  \bibinfo{year}{2018}\natexlab{}.
\newblock \showarticletitle{Systematic Literature Review: {{Self-Regulated
  Learning}} Strategies Using e-{{Learning}} Tools for {{Computer Science}}}.
\newblock \bibinfo{journal}{\emph{Computers \& Education}}
  \bibinfo{volume}{123} (\bibinfo{date}{Aug.} \bibinfo{year}{2018}),
  \bibinfo{pages}{150--163}.
\newblock
\showISSN{03601315}
\href{https://doi.org/10.1016/j.compedu.2018.05.006}{doi:\nolinkurl{10.1016/j.compedu.2018.05.006}}


\bibitem[Giovazolias et~al\mbox{.}(2010)]%
        {Giovazolias2010AssessmentGreekUniversity}
\bibfield{author}{\bibinfo{person}{Theodoros Giovazolias},
  \bibinfo{person}{Sophia Leontopoulou}, {and} \bibinfo{person}{Sophia
  Triliva}.} \bibinfo{year}{2010}\natexlab{}.
\newblock \showarticletitle{Assessment of {{Greek University Students}}'
  {{Counselling Needs}} and {{Attitudes}}: {{An Exploratory Study}}}.
\newblock \bibinfo{journal}{\emph{International Journal for the Advancement of
  Counselling}} \bibinfo{volume}{32}, \bibinfo{number}{2} (\bibinfo{date}{June}
  \bibinfo{year}{2010}), \bibinfo{pages}{101--116}.
\newblock
\showISSN{1573-3246}
\href{https://doi.org/10.1007/s10447-010-9092-2}{doi:\nolinkurl{10.1007/s10447-010-9092-2}}


\bibitem[Glaser and Strauss(1967)]%
        {GlaStr67}
\bibfield{author}{\bibinfo{person}{Barney~G. Glaser} {and}
  \bibinfo{person}{Anselm~L. Strauss}.} \bibinfo{year}{1967}\natexlab{}.
\newblock \bibinfo{booktitle}{\emph{The Discovery of Grounded Theory:
  {{Strategies}} for Qualitative Research}}.
\newblock \bibinfo{publisher}{Aldine Publishing Company}.
\newblock


\bibitem[Gorson and O'Rourke(2020)]%
        {GorOro20}
\bibfield{author}{\bibinfo{person}{Jamie Gorson} {and} \bibinfo{person}{Eleanor
  O'Rourke}.} \bibinfo{year}{2020}\natexlab{}.
\newblock \showarticletitle{Why Do {{CS1}} Students Think They're Bad at
  Programming?}. In \bibinfo{booktitle}{\emph{Proceedings of the 2020 {{ACM}}
  Conference on International Computing Education Research}}.
  \bibinfo{publisher}{ACM}.
\newblock
\href{https://doi.org/10.1145/3372782.3406273}{doi:\nolinkurl{10.1145/3372782.3406273}}


\bibitem[Guba(1981)]%
        {Gub81}
\bibfield{author}{\bibinfo{person}{Egon~G Guba}.}
  \bibinfo{year}{1981}\natexlab{}.
\newblock \showarticletitle{Criteria for Assessing the Trustworthiness of
  Naturalistic Inquiries}.
\newblock \bibinfo{journal}{\emph{ECTJ}} \bibinfo{volume}{29},
  \bibinfo{number}{2} (\bibinfo{year}{1981}), \bibinfo{pages}{75--91}.
\newblock
\href{https://doi.org/10.1007/BF02766777}{doi:\nolinkurl{10.1007/BF02766777}}


\bibitem[Halbert and Nathan(2014)]%
        {HalEtAl14}
\bibfield{author}{\bibinfo{person}{Helen Halbert} {and} \bibinfo{person}{Lisa~P
  Nathan}.} \bibinfo{year}{2014}\natexlab{}.
\newblock \showarticletitle{Designing for Negative Affect and Critical
  Reflection}. In \bibinfo{booktitle}{\emph{{{CHI}}'14 Extended Abstracts on
  Human Factors in Computing Systems}}. \bibinfo{publisher}{ACM},
  \bibinfo{address}{New York, NY}, \bibinfo{pages}{2569--2574}.
\newblock
\href{https://doi.org/10.1145/2559206.2581241}{doi:\nolinkurl{10.1145/2559206.2581241}}


\bibitem[H{\"a}m{\"a}l{\"a}inen and Isom{\"o}tt{\"o}nen(2019)]%
        {Hamalainen2019}
\bibfield{author}{\bibinfo{person}{Ville H{\"a}m{\"a}l{\"a}inen} {and}
  \bibinfo{person}{Ville Isom{\"o}tt{\"o}nen}.}
  \bibinfo{year}{2019}\natexlab{}.
\newblock \showarticletitle{What {{Did CS Students Recognize}} as {{Study
  Difficulties}}?}. In \bibinfo{booktitle}{\emph{2019 {{IEEE Frontiers}} in
  {{Education Conference}} ({{FIE}})}}, Vol.~\bibinfo{volume}{2019-Octob}.
  \bibinfo{publisher}{IEEE}, \bibinfo{pages}{1--9}.
\newblock
\href{https://doi.org/10.1109/FIE43999.2019.9028714}{doi:\nolinkurl{10.1109/FIE43999.2019.9028714}}


\bibitem[Isom{\"o}tt{\"o}nen et~al\mbox{.}(2020)]%
        {Isomottonen2020ExploringStudentsIdentity}
\bibfield{author}{\bibinfo{person}{Ville Isom{\"o}tt{\"o}nen},
  \bibinfo{person}{Ville H{\"a}m{\"a}l{\"a}inen}, \bibinfo{person}{Jennifer
  Clark}, {and} \bibinfo{person}{Vesa Lappalainen}.}
  \bibinfo{year}{2020}\natexlab{}.
\newblock \showarticletitle{Exploring Students' Identity Development from the
  Perspective of Study Difficulties}. In \bibinfo{booktitle}{\emph{2020 {{IEEE
  Frontiers}} in {{Education Conference}} ({{FIE}})}}. \bibinfo{pages}{1--5}.
\newblock
\showISSN{2377-634X}
\href{https://doi.org/10.1109/FIE44824.2020.9273923}{doi:\nolinkurl{10.1109/FIE44824.2020.9273923}}


\bibitem[Isom{\"o}tt{\"o}nen and Nyl{\'e}n(2019)]%
        {IsoNyl19}
\bibfield{author}{\bibinfo{person}{Ville Isom{\"o}tt{\"o}nen} {and}
  \bibinfo{person}{Aletta Nyl{\'e}n}.} \bibinfo{year}{2019}\natexlab{}.
\newblock \showarticletitle{Multiple Authentic Project-Based Experiences and
  Persistent Learning?}. In \bibinfo{booktitle}{\emph{{{IEEE}} Frontiers in
  Education Conference ({{FIE}})}}. \bibinfo{publisher}{IEEE},
  \bibinfo{pages}{1--5}.
\newblock
\href{https://doi.org/10.1109/FIE43999.2019.9028535}{doi:\nolinkurl{10.1109/FIE43999.2019.9028535}}


\bibitem[Isom{\"o}tt{\"o}nen and Tirronen(2016)]%
        {IsoTir16}
\bibfield{author}{\bibinfo{person}{Ville Isom{\"o}tt{\"o}nen} {and}
  \bibinfo{person}{Ville Tirronen}.} \bibinfo{year}{2016}\natexlab{}.
\newblock \showarticletitle{Flipping and Blending---an Action Research Project
  on Improving a Functional Programming Course}.
\newblock \bibinfo{journal}{\emph{ACM Transactions on Computing Education}}
  \bibinfo{volume}{17}, \bibinfo{number}{1}, Article \bibinfo{articleno}{1}
  (\bibinfo{date}{Sept.} \bibinfo{year}{2016}).
\newblock
\href{https://doi.org/10.1145/2934697}{doi:\nolinkurl{10.1145/2934697}}


\bibitem[Kangas et~al\mbox{.}(2017)]%
        {Kangas2017HowFacilitateFreshmen}
\bibfield{author}{\bibinfo{person}{Jari Kangas}, \bibinfo{person}{Elisa
  Rantanen}, {and} \bibinfo{person}{Lauri Kettunen}.}
  \bibinfo{year}{2017}\natexlab{}.
\newblock \showarticletitle{How to Facilitate Freshmen Learning and Support
  Their Transition to a University Study Environment}.
\newblock \bibinfo{journal}{\emph{European Journal of Engineering Education}}
  \bibinfo{volume}{42}, \bibinfo{number}{6} (\bibinfo{date}{Nov.}
  \bibinfo{year}{2017}), \bibinfo{pages}{668--683}.
\newblock
\href{https://doi.org/10.1080/03043797.2016.1214818}{doi:\nolinkurl{10.1080/03043797.2016.1214818}}


\bibitem[Karau and Williams(1993)]%
        {Kar93Social}
\bibfield{author}{\bibinfo{person}{Steven~J. Karau} {and}
  \bibinfo{person}{Kipling~D. Williams}.} \bibinfo{year}{1993}\natexlab{}.
\newblock \showarticletitle{Social Loafing: {{A}} Meta-Analytic Review and
  Theoretical Integration.}
\newblock \bibinfo{journal}{\emph{Journal of Personality and Social
  Psychology}} \bibinfo{volume}{65}, \bibinfo{number}{4} (\bibinfo{date}{Oct.}
  \bibinfo{year}{1993}), \bibinfo{pages}{681--706}.
\newblock
\showISSN{1939-1315, 0022-3514}
\href{https://doi.org/10.1037/0022-3514.65.4.681}{doi:\nolinkurl{10.1037/0022-3514.65.4.681}}


\bibitem[Kierkegaard(1859)]%
        {Kierkegaard1859PointViewMy}
\bibfield{author}{\bibinfo{person}{S{\o}ren Kierkegaard}.}
  \bibinfo{year}{1859}\natexlab{}.
\newblock \bibinfo{booktitle}{\emph{The {{Point}} of {{View}} of {{My Work}} as
  an {{Author}}: {{A Direct Communication}}, {{Report}} to {{History}}}}.
\newblock \bibinfo{publisher}{Princeton University Press},
  \bibinfo{address}{Princeton, New Jersey}.
\newblock
\showISBNx{978-1-4008-3240-8}


\bibitem[Kinnunen and Malmi(2006)]%
        {Kinnunen2006WhyStudentsDrop}
\bibfield{author}{\bibinfo{person}{P{\"a}ivi Kinnunen} {and}
  \bibinfo{person}{Lauri Malmi}.} \bibinfo{year}{2006}\natexlab{}.
\newblock \showarticletitle{Why Students Drop out {{CS1}} Course?}. In
  \bibinfo{booktitle}{\emph{Proceedings of the 2006 International Workshop on
  {{Computing}} Education Research - {{ICER}} '06}}. \bibinfo{publisher}{ACM
  Press}, \bibinfo{address}{New York, New York, USA}, \bibinfo{pages}{97}.
\newblock
\href{https://doi.org/10.1145/1151588.1151604}{doi:\nolinkurl{10.1145/1151588.1151604}}


\bibitem[Lakshminarayanan et~al\mbox{.}(2021)]%
        {LakEtAl21}
\bibfield{author}{\bibinfo{person}{Srinivasan Lakshminarayanan},
  \bibinfo{person}{N.~J. Rao}, {and} \bibinfo{person}{Meghana~G. K.}}
  \bibinfo{year}{2021}\natexlab{}.
\newblock \showarticletitle{Transformative Learning in Mastery-Oriented {{CS0}}
  Course}.
\newblock \bibinfo{journal}{\emph{Higher Education for the Future}}
  \bibinfo{volume}{8}, \bibinfo{number}{2} (\bibinfo{year}{2021}),
  \bibinfo{pages}{162--179}.
\newblock
\href{https://doi.org/10.1177/23476311211007255}{doi:\nolinkurl{10.1177/23476311211007255}}


\bibitem[Lincoln and Guba(1985)]%
        {LinCub85}
\bibfield{author}{\bibinfo{person}{Yvonna~.S. Lincoln} {and}
  \bibinfo{person}{Egon~G. Guba}.} \bibinfo{year}{1985}\natexlab{}.
\newblock \bibinfo{booktitle}{\emph{Naturalistic Inquiry}}.
\newblock \bibinfo{publisher}{Sage Publications}, \bibinfo{address}{Newbury
  Park, CA}.
\newblock


\bibitem[Loksa and Ko(2016)]%
        {Loksa2016RoleSelfRegulationProgramming}
\bibfield{author}{\bibinfo{person}{Dastyni Loksa} {and} \bibinfo{person}{Amy~J.
  Ko}.} \bibinfo{year}{2016}\natexlab{}.
\newblock \showarticletitle{The {{Role}} of {{Self-Regulation}} in
  {{Programming Problem Solving Process}} and {{Success}}}. In
  \bibinfo{booktitle}{\emph{Proceedings of the 2016 {{ACM Conference}} on
  {{International Computing Education Research}}}}. \bibinfo{publisher}{ACM},
  \bibinfo{address}{Melbourne VIC Australia}, \bibinfo{pages}{83--91}.
\newblock
\href{https://doi.org/10.1145/2960310.2960334}{doi:\nolinkurl{10.1145/2960310.2960334}}


\bibitem[MacGregor et~al\mbox{.}(2000)]%
        {MacEtAl00}
\bibfield{author}{\bibinfo{person}{Jean MacGregor}, \bibinfo{person}{Vincent
  Tinto}, {and} \bibinfo{person}{Jerri~Holland Lindblad}.}
  \bibinfo{year}{2000}\natexlab{}.
\newblock \showarticletitle{Assessment of Innovative Efforts: {{Lessons}} from
  the Learning Community Movement}.
\newblock \bibinfo{journal}{\emph{Assessment to promote deep learning: Insight
  from AAHE's}} (\bibinfo{year}{2000}), \bibinfo{pages}{41--48}.
\newblock


\bibitem[Malazita and Resetar(2019)]%
        {MalRes19}
\bibfield{author}{\bibinfo{person}{James~W Malazita} {and}
  \bibinfo{person}{Korryn Resetar}.} \bibinfo{year}{2019}\natexlab{}.
\newblock \showarticletitle{Infrastructures of Abstraction: How Computer
  Science Education Produces Anti-Political Subjects}.
\newblock \bibinfo{journal}{\emph{Digital Creativity}} \bibinfo{volume}{30},
  \bibinfo{number}{4} (\bibinfo{year}{2019}), \bibinfo{pages}{300--312}.
\newblock
\href{https://doi.org/10.1080/14626268.2019.1682616}{doi:\nolinkurl{10.1080/14626268.2019.1682616}}


\bibitem[Marks and Yardley(2004)]%
        {MarYar04}
\bibfield{author}{\bibinfo{person}{D.~F. Marks} {and} \bibinfo{person}{L.
  Yardley}.} \bibinfo{year}{2004}\natexlab{}.
\newblock \showarticletitle{Content and Thematic Analysis}.
\newblock In \bibinfo{booktitle}{\emph{Research Methods for Clinical and Health
  Psychology}}, \bibfield{editor}{\bibinfo{person}{D.~F. Marks} {and}
  \bibinfo{person}{L.~Yardley}} (Eds.). \bibinfo{publisher}{Sage},
  \bibinfo{pages}{56--68}.
\newblock
\href{https://doi.org/10.4135/9781849209793}{doi:\nolinkurl{10.4135/9781849209793}}


\bibitem[Mason(2019)]%
        {Mason2019EvaluationStudySkills}
\bibfield{author}{\bibinfo{person}{Henry~D. Mason}.}
  \bibinfo{year}{2019}\natexlab{}.
\newblock \showarticletitle{Evaluation of a {{Study Skills Intervention
  Programme}}: {{A Mixed Methods Study}}}.
\newblock \bibinfo{journal}{\emph{Africa Education Review}}
  \bibinfo{volume}{16}, \bibinfo{number}{1} (\bibinfo{date}{Jan.}
  \bibinfo{year}{2019}), \bibinfo{pages}{88--105}.
\newblock
\showISSN{1814-6627}
\href{https://doi.org/10.1080/18146627.2016.1241666}{doi:\nolinkurl{10.1080/18146627.2016.1241666}}


\bibitem[Mayer(2004)]%
        {May04}
\bibfield{author}{\bibinfo{person}{Richard~E Mayer}.}
  \bibinfo{year}{2004}\natexlab{}.
\newblock \showarticletitle{Should There Be a Three-Strikes Rule against Pure
  Discovery Learning?}
\newblock \bibinfo{journal}{\emph{American psychologist}} \bibinfo{volume}{59},
  \bibinfo{number}{1} (\bibinfo{year}{2004}), \bibinfo{pages}{14}.
\newblock
\href{https://doi.org/10.1037/0003-066X.59.1.14}{doi:\nolinkurl{10.1037/0003-066X.59.1.14}}


\bibitem[Meehan and Howells(2018)]%
        {Meehan2018SearchFeelingBelonging}
\bibfield{author}{\bibinfo{person}{Catherine Meehan} {and}
  \bibinfo{person}{Kristy Howells}.} \bibinfo{year}{2018}\natexlab{}.
\newblock \showarticletitle{In Search of the Feeling of 'belonging' in Higher
  Education: {{Undergraduate}} Students Transition into Higher Education}.
\newblock \bibinfo{journal}{\emph{Journal of Further and Higher Education}}
  \bibinfo{volume}{43}, \bibinfo{number}{10} (\bibinfo{year}{2018}),
  \bibinfo{pages}{1376--1390}.
\newblock
\href{https://doi.org/10.1080/0309877X.2018.1490702}{doi:\nolinkurl{10.1080/0309877X.2018.1490702}}


\bibitem[Meyer and Land(2005)]%
        {MeyLan05}
\bibfield{author}{\bibinfo{person}{Jan~HF Meyer} {and} \bibinfo{person}{Ray
  Land}.} \bibinfo{year}{2005}\natexlab{}.
\newblock \showarticletitle{Threshold Concepts and Troublesome Knowledge (2):
  {{Epistemological}} Considerations and a Conceptual Framework for Teaching
  and Learning}.
\newblock \bibinfo{journal}{\emph{Higher education}}  \bibinfo{volume}{49}
  (\bibinfo{year}{2005}), \bibinfo{pages}{373--388}.
\newblock
\href{https://doi.org/10.1007/s10734-004-6779-5}{doi:\nolinkurl{10.1007/s10734-004-6779-5}}


\bibitem[Mezirow(1978)]%
        {Mez78}
\bibfield{author}{\bibinfo{person}{Jack Mezirow}.}
  \bibinfo{year}{1978}\natexlab{}.
\newblock \showarticletitle{Perspective Transformation}.
\newblock \bibinfo{journal}{\emph{Adult Education}} \bibinfo{volume}{28},
  \bibinfo{number}{2} (\bibinfo{year}{1978}), \bibinfo{pages}{100--110}.
\newblock
\href{https://doi.org/10.1177/074171367802800202}{doi:\nolinkurl{10.1177/074171367802800202}}


\bibitem[Mezirow(1981)]%
        {Mez81}
\bibfield{author}{\bibinfo{person}{Jack Mezirow}.}
  \bibinfo{year}{1981}\natexlab{}.
\newblock \showarticletitle{A Critical Theory of Adult Learning and Education}.
\newblock \bibinfo{journal}{\emph{Adult Education}} \bibinfo{volume}{32},
  \bibinfo{number}{1} (\bibinfo{year}{1981}), \bibinfo{pages}{3--24}.
\newblock
\href{https://doi.org/10.1177/074171368103200101}{doi:\nolinkurl{10.1177/074171368103200101}}


\bibitem[Miller and Kay(2002)]%
        {MilKay02}
\bibfield{author}{\bibinfo{person}{Amanda Miller} {and} \bibinfo{person}{Judy
  Kay}.} \bibinfo{year}{2002}\natexlab{}.
\newblock \showarticletitle{A Mentor Program in {{CS1}}}.
\newblock \bibinfo{journal}{\emph{ACM SIGCSE Bulletin}} \bibinfo{volume}{34},
  \bibinfo{number}{3} (\bibinfo{date}{Sept.} \bibinfo{year}{2002}),
  \bibinfo{pages}{9--13}.
\newblock
\showISSN{0097-8418}
\href{https://doi.org/10.1145/637610.544420}{doi:\nolinkurl{10.1145/637610.544420}}


\bibitem[Minnes et~al\mbox{.}(2018)]%
        {Minnes2018LightweightTechniquesSupport}
\bibfield{author}{\bibinfo{person}{Mia Minnes}, \bibinfo{person}{Christine
  Alvarado}, {and} \bibinfo{person}{Leo Porter}.}
  \bibinfo{year}{2018}\natexlab{}.
\newblock \showarticletitle{Lightweight {{Techniques}} to {{Support Students}}
  in {{Large Classes}}}. In \bibinfo{booktitle}{\emph{Proceedings of the 49th
  {{ACM Technical Symposium}} on {{Computer Science Education}}}}
  \emph{(\bibinfo{series}{{{SIGCSE}} '18})}. \bibinfo{publisher}{Association
  for Computing Machinery}, \bibinfo{address}{New York, NY, USA},
  \bibinfo{pages}{122--127}.
\newblock
\href{https://doi.org/10.1145/3159450.3159601}{doi:\nolinkurl{10.1145/3159450.3159601}}


\bibitem[Mirza et~al\mbox{.}(2019)]%
        {Mirza2019UndergraduateTeachingAssistants}
\bibfield{author}{\bibinfo{person}{Diba Mirza}, \bibinfo{person}{Phillip~T.
  Conrad}, \bibinfo{person}{Christian Lloyd}, \bibinfo{person}{Ziad Matni},
  {and} \bibinfo{person}{Arthur Gatin}.} \bibinfo{year}{2019}\natexlab{}.
\newblock \showarticletitle{Undergraduate {{Teaching Assistants}} in {{Computer
  Science}}: {{A Systematic Literature Review}}}. In
  \bibinfo{booktitle}{\emph{Proceedings of the 2019 {{ACM Conference}} on
  {{International Computing Education Research}}}}. \bibinfo{publisher}{ACM},
  \bibinfo{address}{Toronto ON Canada}, \bibinfo{pages}{31--40}.
\newblock
\href{https://doi.org/10.1145/3291279.3339422}{doi:\nolinkurl{10.1145/3291279.3339422}}


\bibitem[Morris et~al\mbox{.}(2012)]%
        {MorEtAl12}
\bibfield{author}{\bibinfo{person}{Arlene~H Morris}, \bibinfo{person}{Debbie~R
  Faulk}, {and} \bibinfo{person}{Michelle~A Schutt}.}
  \bibinfo{year}{2012}\natexlab{}.
\newblock \showarticletitle{Self-Regulation through Transformative Learning}.
\newblock In \bibinfo{booktitle}{\emph{Transformative Learning in Nursing: A
  Guide for Nurse Educators}}, \bibfield{editor}{\bibinfo{person}{Arlene~H
  Morris} {and} \bibinfo{person}{Debbie~R Faulk}} (Eds.).
  \bibinfo{publisher}{Springer Publishing Company}, \bibinfo{pages}{168}.
\newblock


\bibitem[Mostr{\"o}m et~al\mbox{.}(2008)]%
        {MosEtAl08}
\bibfield{author}{\bibinfo{person}{Jan~Erik Mostr{\"o}m},
  \bibinfo{person}{Jonas Boustedt}, \bibinfo{person}{Anna Eckerdal},
  \bibinfo{person}{Robert McCartney}, \bibinfo{person}{Kate Sanders},
  \bibinfo{person}{Lynda Thomas}, {and} \bibinfo{person}{Carol Zander}.}
  \bibinfo{year}{2008}\natexlab{}.
\newblock \showarticletitle{Concrete Examples of Abstraction as Manifested in
  Students' Transformative Experiences}. In
  \bibinfo{booktitle}{\emph{Proceedings of the Fourth International Workshop on
  Computing Education Research}}. \bibinfo{publisher}{ACM},
  \bibinfo{pages}{125--136}.
\newblock
\href{https://doi.org/10.1145/1404520.1404533}{doi:\nolinkurl{10.1145/1404520.1404533}}


\bibitem[Peavy(1998)]%
        {Pea98}
\bibfield{author}{\bibinfo{person}{R~Vance Peavy}.}
  \bibinfo{year}{1998}\natexlab{}.
\newblock \bibinfo{booktitle}{\emph{Sociodynamic Counselling: {{A}}
  Constructivist Perspective}}.
\newblock \bibinfo{publisher}{Trafford}, \bibinfo{address}{Victoria, BC,
  Canada}.
\newblock


\bibitem[Perin(2011)]%
        {Perin2011FacilitatingStudentLearning}
\bibfield{author}{\bibinfo{person}{Dolores Perin}.}
  \bibinfo{year}{2011}\natexlab{}.
\newblock \showarticletitle{Facilitating {{Student Learning Through
  Contextualization}}: {{A Review}} of {{Evidence}}}.
\newblock \bibinfo{journal}{\emph{Community College Review}}
  \bibinfo{volume}{39}, \bibinfo{number}{3} (\bibinfo{date}{July}
  \bibinfo{year}{2011}), \bibinfo{pages}{268--295}.
\newblock
\showISSN{0091-5521}
\href{https://doi.org/10.1177/0091552111416227}{doi:\nolinkurl{10.1177/0091552111416227}}


\bibitem[Perry et~al\mbox{.}(2001)]%
        {Perry2001AcademicControlAction}
\bibfield{author}{\bibinfo{person}{Raymond~P. Perry}, \bibinfo{person}{Steven
  Hladkyj}, \bibinfo{person}{Sarah~T. Pelletier}, {and}
  \bibinfo{person}{Reinhard~H. Pekrun}.} \bibinfo{year}{2001}\natexlab{}.
\newblock \showarticletitle{Academic Control and Action Control in the
  Achievement of College Students: {{A}} Longitudinal Field Study}.
\newblock \bibinfo{journal}{\emph{Journal of Educational Psychology}}
  \bibinfo{volume}{93}, \bibinfo{number}{4} (\bibinfo{year}{2001}),
  \bibinfo{pages}{776--789}.
\newblock
\showISSN{00220663}
\href{https://doi.org/10.1037/0022-0663.93.4.776}{doi:\nolinkurl{10.1037/0022-0663.93.4.776}}


\bibitem[Petersen et~al\mbox{.}(2016)]%
        {Petersen2016RevisitingWhyStudents}
\bibfield{author}{\bibinfo{person}{Andrew Petersen}, \bibinfo{person}{Michelle
  Craig}, \bibinfo{person}{Jennifer Campbell}, {and} \bibinfo{person}{Anya
  Tafliovich}.} \bibinfo{year}{2016}\natexlab{}.
\newblock \showarticletitle{Revisiting Why Students Drop {{CS1}}}. In
  \bibinfo{booktitle}{\emph{Proceedings of the 16th {{Koli Calling
  International Conference}} on {{Computing Education Research}}}}.
  \bibinfo{publisher}{ACM}, \bibinfo{address}{New York, NY, USA},
  \bibinfo{pages}{71--80}.
\newblock
\href{https://doi.org/10.1145/2999541.2999552}{doi:\nolinkurl{10.1145/2999541.2999552}}


\bibitem[Pintrich(2000)]%
        {Pintrich2000RoleGoalOrientation}
\bibfield{author}{\bibinfo{person}{Paul~R. Pintrich}.}
  \bibinfo{year}{2000}\natexlab{}.
\newblock \showarticletitle{The {{Role}} of {{Goal Orientation}} in
  {{Self-Regulated Learning}}}.
\newblock In \bibinfo{booktitle}{\emph{Handbook of {{Self-Regulation}}}},
  \bibfield{editor}{\bibinfo{person}{Monique Boekaerts},
  \bibinfo{person}{Paul~R. Pintrich}, {and} \bibinfo{person}{Moshe Zeidner}}
  (Eds.). \bibinfo{publisher}{Elsevier}, \bibinfo{pages}{451--502}.
\newblock
\href{https://doi.org/10.1016/b978-012109890-2/50043-3}{doi:\nolinkurl{10.1016/b978-012109890-2/50043-3}}


\bibitem[{Pon-Barry} et~al\mbox{.}(2017)]%
        {PonEtAl17}
\bibfield{author}{\bibinfo{person}{Heather {Pon-Barry}}, \bibinfo{person}{Becky
  Wai-Ling Packard}, {and} \bibinfo{person}{Audrey St.~John}.}
  \bibinfo{year}{2017}\natexlab{}.
\newblock \showarticletitle{Expanding Capacity and Promoting Inclusion in
  Introductory Computer Science: A Focus on near-{{Peer}} Mentor Preparation
  and Code Review}.
\newblock \bibinfo{journal}{\emph{Computer Science Education}}
  \bibinfo{volume}{27}, \bibinfo{number}{1} (\bibinfo{date}{Jan.}
  \bibinfo{year}{2017}), \bibinfo{pages}{54--77}.
\newblock
\showISSN{0899-3408}
\href{https://doi.org/10.1080/08993408.2017.1333270}{doi:\nolinkurl{10.1080/08993408.2017.1333270}}


\bibitem[Prather et~al\mbox{.}(2020)]%
        {Prather2020WhatWeThinka}
\bibfield{author}{\bibinfo{person}{James Prather}, \bibinfo{person}{Brett~A.
  Becker}, \bibinfo{person}{Michelle Craig}, \bibinfo{person}{Paul Denny},
  \bibinfo{person}{Dastyni Loksa}, {and} \bibinfo{person}{Lauren Margulieux}.}
  \bibinfo{year}{2020}\natexlab{}.
\newblock \showarticletitle{What {{Do We Think We Think We Are Doing}}?:
  {{Metacognition}} and {{Self-Regulation}} in {{Programming}}}. In
  \bibinfo{booktitle}{\emph{Proceedings of the 2020 {{ACM Conference}} on
  {{International Computing Education Research}}}}. \bibinfo{publisher}{ACM},
  \bibinfo{address}{Virtual Event New Zealand}, \bibinfo{pages}{2--13}.
\newblock
\href{https://doi.org/10.1145/3372782.3406263}{doi:\nolinkurl{10.1145/3372782.3406263}}


\bibitem[Rogerson and Scott(2010)]%
        {RogSco10}
\bibfield{author}{\bibinfo{person}{Christine Rogerson} {and}
  \bibinfo{person}{Elsje Scott}.} \bibinfo{year}{2010}\natexlab{}.
\newblock \showarticletitle{The Fear Factor: {{How}} It Affects Students
  Learning to Program in a Tertiary Environment}.
\newblock \bibinfo{journal}{\emph{Journal of Information Technology Education:
  Research}}  \bibinfo{volume}{9} (\bibinfo{year}{2010}),
  \bibinfo{pages}{147--171}.
\newblock


\bibitem[Sandelowski(2010)]%
        {San10}
\bibfield{author}{\bibinfo{person}{Margarete Sandelowski}.}
  \bibinfo{year}{2010}\natexlab{}.
\newblock \showarticletitle{What's in a Name? {{Qualitative}} Description
  Revisited}.
\newblock \bibinfo{journal}{\emph{Research in nursing \& health}}
  \bibinfo{volume}{33}, \bibinfo{number}{1} (\bibinfo{year}{2010}),
  \bibinfo{pages}{77--84}.
\newblock
\href{https://doi.org/10.1002/nur.20362}{doi:\nolinkurl{10.1002/nur.20362}}


\bibitem[{Santillan-Rosas} and {Heredia-Escorza}(2020)]%
        {SanEtal20}
\bibfield{author}{\bibinfo{person}{Irais~Monserrat {Santillan-Rosas}} {and}
  \bibinfo{person}{Yolanda {Heredia-Escorza}}.}
  \bibinfo{year}{2020}\natexlab{}.
\newblock \showarticletitle{Empowering Women's Digital Literacy with
  Transformative Learning: {{Reducing}} the Gap in the {{T}} of {{STEM}}}. In
  \bibinfo{booktitle}{\emph{Eighth International Conference on Technological
  Ecosystems for Enhancing Multiculturality}}. \bibinfo{publisher}{ACM},
  \bibinfo{address}{New York, NY}, \bibinfo{pages}{182--186}.
\newblock
\href{https://doi.org/10.1145/3434780.3436684}{doi:\nolinkurl{10.1145/3434780.3436684}}


\bibitem[Schunk(2005)]%
        {Schunk2005}
\bibfield{author}{\bibinfo{person}{Dale~H Schunk}.}
  \bibinfo{year}{2005}\natexlab{}.
\newblock \showarticletitle{Self-{{Regulated Learning}}: {{The Educational
  Legacy}} of {{Paul R}}. {{Pintrich}}}.
\newblock \bibinfo{journal}{\emph{Educational Psychologist}}
  \bibinfo{volume}{40}, \bibinfo{number}{2} (\bibinfo{date}{June}
  \bibinfo{year}{2005}), \bibinfo{pages}{85--94}.
\newblock
\showISSN{0046-1520}
\href{https://doi.org/10.1207/s15326985ep4002_3}{doi:\nolinkurl{10.1207/s15326985ep4002_3}}


\bibitem[Schwartzman(2007)]%
        {Sch07}
\bibfield{author}{\bibinfo{person}{Leslie Schwartzman}.}
  \bibinfo{year}{2007}\natexlab{}.
\newblock \showarticletitle{Student Transformative Learning in Software
  Engineering and Design: Discontinuity (Pre) Serves Meaning}. In
  \bibinfo{booktitle}{\emph{Proceedings of the Seventh Baltic Sea Conference on
  Computing Education Research-Volume 88 (Koli Calling)}}.
  \bibinfo{pages}{97--108}.
\newblock


\bibitem[Steinhorst et~al\mbox{.}(2020)]%
        {Steinhorst2020RevisitingSelfEfficacyIntroductory}
\bibfield{author}{\bibinfo{person}{Phil Steinhorst}, \bibinfo{person}{Andrew
  Petersen}, {and} \bibinfo{person}{Jan Vahrenhold}.}
  \bibinfo{year}{2020}\natexlab{}.
\newblock \showarticletitle{Revisiting {{Self-Efficacy}} in {{Introductory
  Programming}}}. In \bibinfo{booktitle}{\emph{Proceedings of the 2020 {{ACM
  Conference}} on {{International Computing Education Research}}}}.
  \bibinfo{publisher}{ACM}, \bibinfo{address}{New York, NY, USA},
  \bibinfo{pages}{158--169}.
\newblock
\href{https://doi.org/10.1145/3372782.3406281}{doi:\nolinkurl{10.1145/3372782.3406281}}


\bibitem[Thi Van~Pham et~al\mbox{.}(2021)]%
        {ThiVanPham2021StudentSupportServices}
\bibfield{author}{\bibinfo{person}{Anh Thi Van~Pham}, \bibinfo{person}{Nam
  Van~Kieu}, {and} \bibinfo{person}{Thao Thi Thu~Vu}.}
  \bibinfo{year}{2021}\natexlab{}.
\newblock \showarticletitle{Student {{Support Services}} in an {{Online
  Learning Environment}}}. In \bibinfo{booktitle}{\emph{2021 5th
  {{International Conference}} on {{E-Society}}, {{E-Education}} and
  {{E-Technology}}}} \emph{(\bibinfo{series}{{{ICSET}} 2021})}.
  \bibinfo{publisher}{Association for Computing Machinery},
  \bibinfo{address}{New York, NY, USA}, \bibinfo{pages}{14--19}.
\newblock
\href{https://doi.org/10.1145/3485768.3485801}{doi:\nolinkurl{10.1145/3485768.3485801}}


\bibitem[Thomas(2009)]%
        {Tho09}
\bibfield{author}{\bibinfo{person}{Ian Thomas}.}
  \bibinfo{year}{2009}\natexlab{}.
\newblock \showarticletitle{Critical Thinking, Transformative Learning,
  Sustainable Education, and Problem-Based Learning in Universities}.
\newblock \bibinfo{journal}{\emph{Journal of Transformative Education}}
  \bibinfo{volume}{7}, \bibinfo{number}{3} (\bibinfo{year}{2009}),
  \bibinfo{pages}{245--264}.
\newblock
\href{https://doi.org/10.1177/1541344610385753}{doi:\nolinkurl{10.1177/1541344610385753}}


\bibitem[Turner et~al\mbox{.}(2017)]%
        {Turner2017EasingTransitionFirst}
\bibfield{author}{\bibinfo{person}{Rebecca Turner}, \bibinfo{person}{David
  Morrison}, \bibinfo{person}{Debby Cotton}, \bibinfo{person}{Samantha Child},
  \bibinfo{person}{Sebastian Stevens}, \bibinfo{person}{Patricia Nash}, {and}
  \bibinfo{person}{Pauline Kneale}.} \bibinfo{year}{2017}\natexlab{}.
\newblock \showarticletitle{Easing the Transition of First Year Undergraduates
  through an Immersive Induction Module}.
\newblock \bibinfo{journal}{\emph{Teaching in Higher Education}}
  \bibinfo{volume}{22}, \bibinfo{number}{7} (\bibinfo{date}{Oct.}
  \bibinfo{year}{2017}), \bibinfo{pages}{805--821}.
\newblock
\href{https://doi.org/10.1080/13562517.2017.1301906}{doi:\nolinkurl{10.1080/13562517.2017.1301906}}


\bibitem[Vaismoradi et~al\mbox{.}(2013)]%
        {VaiEtAl13}
\bibfield{author}{\bibinfo{person}{Mojtaba Vaismoradi},
  \bibinfo{person}{Hannele Turunen}, {and} \bibinfo{person}{Terese Bondas}.}
  \bibinfo{year}{2013}\natexlab{}.
\newblock \showarticletitle{Content Analysis and Thematic Analysis:
  {{Implications}} for Conducting a Qualitative Descriptive Study}.
\newblock \bibinfo{journal}{\emph{Nursing \& health sciences}}
  \bibinfo{volume}{15}, \bibinfo{number}{3} (\bibinfo{year}{2013}),
  \bibinfo{pages}{398--405}.
\newblock
\href{https://doi.org/10.1111/nhs.12048}{doi:\nolinkurl{10.1111/nhs.12048}}


\bibitem[{van Dam}(2018)]%
        {vanDam2018ReflectionsIntroductoryCS}
\bibfield{author}{\bibinfo{person}{Andries {van Dam}}.}
  \bibinfo{year}{2018}\natexlab{}.
\newblock \showarticletitle{Reflections on an Introductory {{CS}} Course,
  {{CS15}}, at {{Brown University}}}.
\newblock \bibinfo{journal}{\emph{ACM Inroads}} \bibinfo{volume}{9},
  \bibinfo{number}{4} (\bibinfo{date}{Nov.} \bibinfo{year}{2018}),
  \bibinfo{pages}{58--62}.
\newblock
\showISSN{2153-2184}
\href{https://doi.org/10.1145/3284639}{doi:\nolinkurl{10.1145/3284639}}


\bibitem[Ventura~Jr(2005)]%
        {Ven05}
\bibfield{author}{\bibinfo{person}{Philip~R Ventura~Jr}.}
  \bibinfo{year}{2005}\natexlab{}.
\newblock \showarticletitle{Identifying Predictors of Success for an
  Objects-First {{CS1}}}.
\newblock \bibinfo{journal}{\emph{Computer Science Education}}
  \bibinfo{volume}{15}, \bibinfo{number}{3} (\bibinfo{year}{2005}).
\newblock
\href{https://doi.org/10.1080/08993400500224419}{doi:\nolinkurl{10.1080/08993400500224419}}


\bibitem[Vihavainen et~al\mbox{.}(2011)]%
        {VihEtAl11}
\bibfield{author}{\bibinfo{person}{Arto Vihavainen}, \bibinfo{person}{Matti
  Paksula}, {and} \bibinfo{person}{Matti Luukkainen}.}
  \bibinfo{year}{2011}\natexlab{}.
\newblock \showarticletitle{Extreme Apprenticeship Method in Teaching
  Programming for Beginners}. In \bibinfo{booktitle}{\emph{Proceedings of the
  42nd {{ACM}} Technical Symposium on {{Computer}} Science Education}}.
  \bibinfo{publisher}{ACM}, \bibinfo{address}{New York, NY},
  \bibinfo{pages}{93--98}.
\newblock
\href{https://doi.org/10.1145/1953163.1953196}{doi:\nolinkurl{10.1145/1953163.1953196}}


\bibitem[Wilson and Shrock(2001)]%
        {WilSch01}
\bibfield{author}{\bibinfo{person}{Brenda~Cantwell Wilson} {and}
  \bibinfo{person}{Sharon Shrock}.} \bibinfo{year}{2001}\natexlab{}.
\newblock \showarticletitle{Contributing to Success in an Introductory Computer
  Science Course: A Study of Twelve Factors}.
\newblock \bibinfo{journal}{\emph{Acm sigcse bulletin}} \bibinfo{volume}{33},
  \bibinfo{number}{1} (\bibinfo{year}{2001}), \bibinfo{pages}{184--188}.
\newblock
\href{https://doi.org/10.1145/366413.364581}{doi:\nolinkurl{10.1145/366413.364581}}


\bibitem[Zimmerman(2008)]%
        {Zimmerman2008InvestigatingSelfRegulationMotivation}
\bibfield{author}{\bibinfo{person}{Barry~J. Zimmerman}.}
  \bibinfo{year}{2008}\natexlab{}.
\newblock \showarticletitle{Investigating {{Self-Regulation}} and
  {{Motivation}}: {{Historical Background}}, {{Methodological Developments}},
  and {{Future Prospects}}}.
\newblock \bibinfo{journal}{\emph{American Educational Research Journal}}
  \bibinfo{volume}{45}, \bibinfo{number}{1} (\bibinfo{date}{March}
  \bibinfo{year}{2008}), \bibinfo{pages}{166--183}.
\newblock
\showISSN{0002-8312}
\href{https://doi.org/10.3102/0002831207312909}{doi:\nolinkurl{10.3102/0002831207312909}}


\end{thebibliography}

\appendix
\section{Illustration of process} \label{app:illustration}

The excerpt below illustrates interviews for the reader. This example (usr15) refers to developed self-regulation which the student links with previous study difficulties. \\

\noindent The student first describes how they manage tasks:

\begin{quote}
[Student:] Oh, and then I had a phone where I put it.
Because that's where you can put those checkboxes in the notebook. It is satisfying to get a kind of mental sense of achievement when you can click on them, and then it goes grey from there. Kind of like extra motivation.

Researcher 2: Do you use that in other courses?
\end{quote}
\noindent Below, student passingly refers to an important other who had contributed to starting a routine.
\begin{quote}
[Student:] I started this some time a year ago. On a few courses I did the same thing, that I put the concrete tasks divided into small pieces [into it].
It was because of that, when I first time talked to [a staff member] I heard this recommendation. It's this self-development. When through this routine one has noticed that it's easier [as you] organize visually for it. For me it was important that courses are not left uncompleted. That's it. That's the motivator.

[...]

[Researcher 1:] Okay. How did you do the weekly exercises? ... were you at the weekly lab sessions with someone? or did you work mainly alone? Or was the TA potentially the only contact person?

[Student:] I talked with a friend. But not about any exercises but just about the course generally. [The friend] had already taken the course in the spring 2020, when I started. So I talked to him, but not regarding any real [programming] matter. It's like [the friend] doesn't remember anything. I wouldn't have 
got any help from him or anything. Yeah, I didn't really talk [much with others]. In other courses, I have made a couple of friends, but here I didn't really try.

[Researcher 1]: So you have studied this course quite independently.
\end{quote}
\noindent Below, student refers to working together as a potentially hindering factor.
\begin{quote} 
Student: Yes, because I felt that one had to do most of the work at home anyway.
I mean, at home I did mostly the exercises. I need to be able to concentrate fully, without anyone talking. It's often so inefficient to ask questions from someone else. That you have to be so ready and present it in the right way. The question... if you want that some other student can help, as they don't have the instructor's perspective and expertise.

[Researcher 1:] Do you think it would be... If there was someone with whom you could discuss programming, would they be more approachable than TAs? Or is it the TA to whom you would prefer to turn if you had a problem?

[Student:] Yes, exactly these student TAs. I didn't talk to [lecturer] myself except in Zoom [...] but I favor these fresh student TAs the most. 

[Researcher 2:] In summary, you appear to have self-regulation skills, which we tend to talk here on the teaching side. You seem to want to be aware of your study situation. That's a very good thing.
\end{quote}
\noindent Finally, as quoted in Section \ref{subsubusec:disruptive-experiences}, the student explicitly noted what the development of self-regulation had personally indicated:
\begin{quote}
[The learning improved routine] evolved through frustration and wasting time. This [personal] development has come with a bitter and high cost.
\end{quote}

\noindent Earlier in the interview, the student referred to considerable difficulties when taking this course earlier. 

\begin{figure}[h]
\centering
\includegraphics[width=10cm]{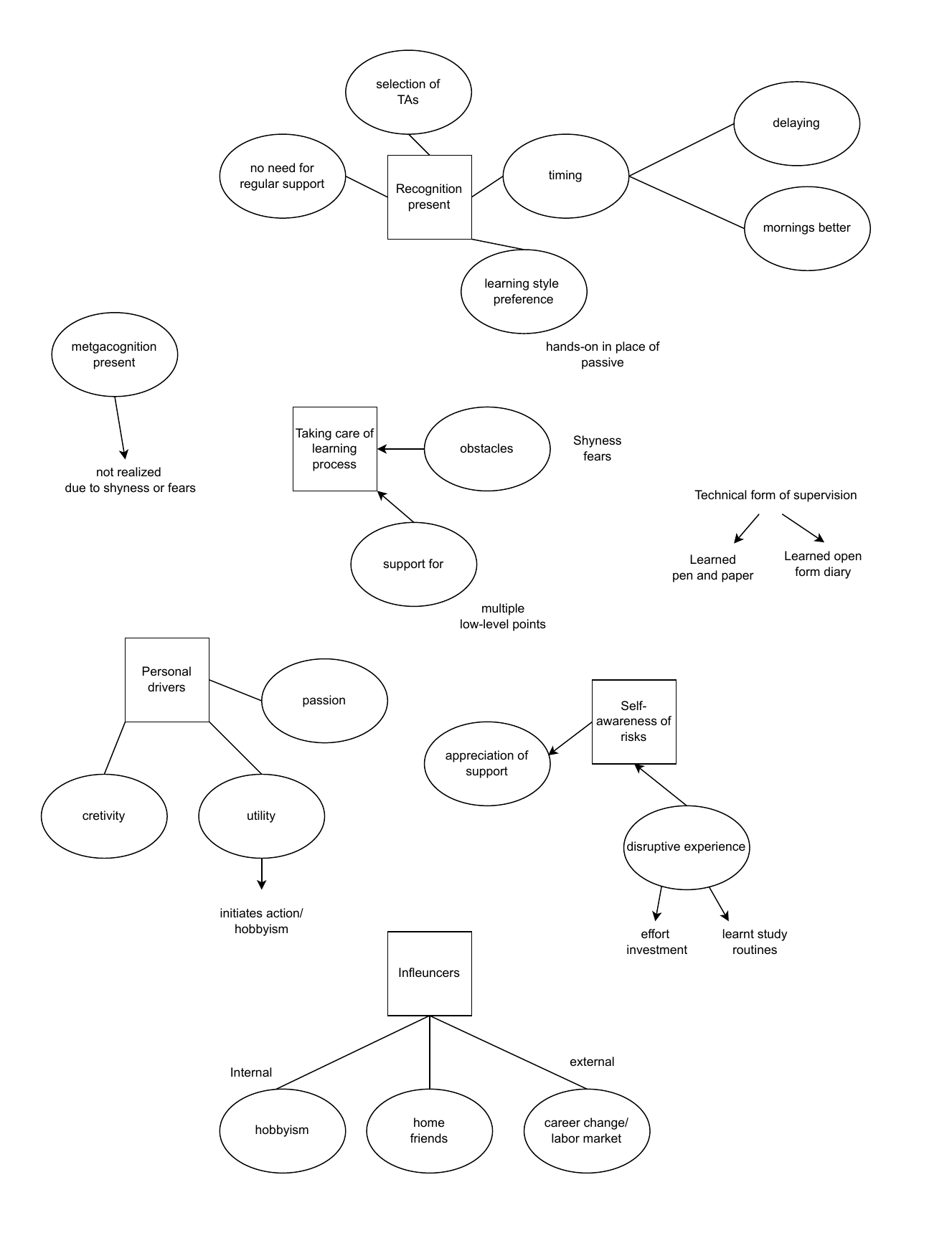}
\caption{Suggestive connections in the data drafted by the first author before collaborative work}
\label{fig:suggestive-connections}
\end{figure}

\end{document}